\documentclass{aa}  

\usepackage{graphicx}
\usepackage{txfonts}
\usepackage{lipsum}
\usepackage{subcaption}         
\usepackage{lscape}             
\usepackage{placeins}           

\usepackage[breaklinks,colorlinks,citecolor=blue,pdfa=true]{hyperref}
\usepackage{color}

\newcommand{\GG}[1]{}

\def\simi   {$\sim$\,}

\def\kms   {km s$^{-1}$ }

\def\mr    {m$_{\rm r}$ }

\begin{document}

   \title{SAGAN-VI: When Jets Meet Filaments -- Environmental Imprints on the Growth of Giant Radio Galaxies}



\author{Mousumi Mahato\inst{1}\thanks{E-mail: mousumi.mahato@ut.ee}
\and Elmo Tempel\inst{1}
\and Shishir Sankhyayan\inst{1}
\and Pratik Dabhade\inst{2}
\and Kshitij Chavan\inst{3}
}
 
\authorrunning{Mahato et al.}
\institute{$^{1}$Tartu Observatory, University of Tartu, Observatooriumi~1, 61602 T\~oravere, Estonia\\ 
$^{2}$Astrophysics Division, National Centre for Nuclear Research, Pasteura 7, 02-093 Warsaw, Poland\\
$^{3}$Inter-University Centre for Astronomy and Astrophysics (IUCAA), Pune 411007, India\\}


 
  \abstract
   {Giant radio galaxies (GRGs) represent the largest individual astrophysical structures, rivalling galaxy clusters in physical extent. Understanding how they attain such scales demands examining their large-scale cosmic surroundings, particularly the under-explored filament environment.} 
   { We quantify the three-dimensional (3D) distance of GRGs from the nearest filament spine; test how this distance correlates with their growth and formation of different morphological classes; assess whether their radio jets exhibit preferred orientations relative to filament axes; and examine how filament anisotropy from spine–to-periphery modulates radio morphology.}
   {We employed a filament catalogue from the Sloan Digital Sky Survey (SDSS) together with the largest GRG catalogue currently available. For each source, we measured the comoving distance to the nearest filament spine, the projected jet-spine orientation angle, and quantified lobe asymmetry via the arm-length ratio (ALR). These metrics trace proximity, directionality, 
   and the impact of filamentary environment on morphology. We then compared GRGs with a control sample of small radio galaxies (SRGs) to constrain the environmental factors that regulate the attainment of giant sizes. We validated the robustness of our results via bootstrap resampling and non-parametric statistical tests.}
   {Our results show that GRGs and SRGs have similar filament occupancy. By contrast, GRGs preferentially display larger alignment angles relative to filament spines, while SRG orientations are consistent with a random distribution. GRGs further show enhanced morphological asymmetry, reflected in lower ALR values than SRGs.}
   {Attainment of giant sizes is not governed by proximity to filaments; rather, it correlates with jet-filament alignment. GRGs are preferentially oriented at large angles to filament spines, consistent with the propagation of jets through lower-density void-facing channels that minimise environmental resistance. Consistently, lobe asymmetry mirrors this alignment, indicating that GRGs experience steeper transverse spine-to-void differential ram-pressure gradients along their paths.} 


   \maketitle

\section{Introduction} \label{sec:intro}

Powerful radio galaxies (RGs) drive magnetised, collimated, relativistic jets \citep{Blandford2019} into the intergalactic medium (IGM), making them exquisitely sensitive, spatially extended probes of galaxy-environment coupling from tens of kiloparsecs (kpc) to several megaparsecs (Mpc). In the canonical Fanaroff-Riley framework, collimated, relativistic FRII jets terminate in compact hotspots and inflate synchrotron-bright lobes, whereas jets that decelerate more gradually produce edge-darkened, diffuse FRI structures \citep{FR74}.
RGs can be further classified based on their projected linear sizes: giant radio galaxies (GRGs) are systems with sizes exceeding 0.7 Mpc, and small radio galaxies (SRGs) have sizes below 0.7 Mpc. GRGs \citep[for review, see;][]{Dabhade2023} with their extreme sizes mark the ultimate stage of growth of RGs, and therefore, are natural laboratories for testing how ambient density, pressure, and magnetisation modulate jet propagation and lobe evolution.  Born in the host galaxy, the GRG jets propagate through the IGM, and eventually emerge onto the scales of the cosmic web; en route, they encounter a variety of environments whose influences are imprinted in the form of lobe-size and flux-density asymmetries, altered collimation and depolarisation patterns, as well as in signatures of episodic nuclear activity \citep[e.g.,][]{sagan5}. Decoding these diagnostics will help identify the physical conditions that permit their growth to Mpc scales.

Since the pioneering discovery of GRGs 3C236 and DA240 by \citet{Willis1974} to the recent and largest one, \textit{Porphyrion} ($\sim$\,7 Mpc) by \citet{Oei2024}, the radio telescopes, with improving resolution, sensitivity and sky coverage, have catalogued nearly 12000 GRGs \citep[e.g.,][]{Mostert2024} till now. Yet, despite this progress, GRGs appear to be relatively rarer compared to SRGs. The extraordinary sizes of GRGs are generally attributed to three main scenarios. First, they may be powered by exceptionally powerful and long-lived AGN activity, driving jets to large distances \citep[e.g.,][]{Gopal1989, wita-grg-agn, DabhadeSAGAN20}. Second, their growth could result from recurrent or episodic jet activity, where multiple cycles of outbursts extend the lobes over time \citep{sagan5}. Third, GRGs are often found in relatively low-density environments, where the reduced external pressure allows jets to propagate farther without significant disruption \citep[e.g.,][]{mack98, Malarecki2015, Stuardi2020}.

Both \citet{Casadei2024} and \citet{Massaro2019} show that RGs, regardless of morphology (FRI or FRII) or optical excitation class (low or high), occupy a wide variety of environments ranging from poor groups to rich clusters, with no statistically significant difference in environmental richness between these categories. Their results consistently indicate that radio morphology and excitation type are not primarily governed by the large-scale environment at low redshift. The main distinction lies in their methods: \citet{Casadei2024} used photometric red-sequence galaxy counts within 0.5 Mpc of each radio galaxy, probing the inner cluster regions, while \citet{Massaro2019} used spectroscopic neighbour counts within 2 Mpc, characterising the broader group and cluster environments. Despite these differences in scale and sample selection (bright 3CR sources in comparison with the more homogeneous FRI and FRII samples), both studies reach the same conclusion that the surrounding environment plays only a secondary role, while intrinsic factors such as host galaxy properties and AGN power are the dominant influences on radio morphology and activity. Similarly, focusing specifically on GRGs, \citet{Lan2021} examined 110 GRGs and found no correlation between their sizes and local galaxy density or distance to large-scale filaments. Their results show that GRGs inhabit environments comparable to those of SRGs, suggesting that the growth of radio structures is driven primarily by intrinsic factors rather than external environmental conditions.

Studies by \citet{PDLOTSS, DabhadeSAGAN20} first showed that around 10-20$\%$ of GRGs are located at the centres of galaxy clusters, posing a challenge to the primary hypothesis that GRGs preferentially grow in low-density environments. A recent localisation of luminous GRGs within Bayesian reconstructions of the cosmic web \citep{Oei2024env} sharpens this picture by revealing that giants prefer denser structures akin to RGs in general, consistent with a radio-luminosity-density relation in which powerful jets preferentially occur in dense environments. Based on their analysis, around $\sim$\,24$\%$ of giants inhabit galaxy clusters, and for $\sim$\,60$\%$ of GRGs, the most probable environment is filament with negligible void occupancy. Complementary work by \citet{Sankhyayan2024} finds that $\sim$\,24$\%$ of GRGs are cluster-associated, with $\sim$\,5$\%$ of them in superclusters. They also highlight that the largest GRGs ($\gtrsim$\,3 Mpc) grow in sparse environments. Moreover, case studies of the largest giants: \textit{Alcyoneus} ($\sim$\,5~Mpc) and \textit{Porphyrion} ($\sim$\,7~Mpc) reveal that they are likely growing within cosmic filaments \citep{Oei2022, Oei2024}, yet the lobes of \textit{Porphyrion} are penetrating cosmic voids with a high probability. \textit{Porphyrion}, in particular, exemplifies how powerful radio outflows can arise from dense filaments and channel energy far into low-density volumes.

As part of the Search and analysis of giant radio galaxies with associated nuclei
(SAGAN) project\footnote{\url{https://sites.google.com/site/anantasakyatta/sagan}}
\citep{DabhadeSAGAN20}, we aim to investigate how the large-scale, filamentary structure of the cosmic web influences the growth and evolution of GRGs. Specifically, we examine the role of the less well-characterised cosmic web filaments in shaping their morphology and evolutionary pathways. These filaments are the Universe's intricate, grand architecture, woven from galaxies, gas, and dark matter, stretching across several megaparsecs. Serving as cosmic bridges between galaxy groups and clusters, they channel gas into galaxies, fuel star formation and drive galaxy evolution. The advent of wide, highly sensitive radio surveys together with deep optical data now makes it possible to investigate environments in unprecedented detail, allowing us to precisely localise radio galaxies within cosmic filaments, probe their interactions, and address key questions such as: 
\begin{itemize}
    \item whether there is any difference in occupancy of GRGs and SRGs in filaments;
    \item how three-dimensional (3D) distance from filaments relates to morphology and projected linear sizes;
    \item whether radio-jet axes exhibit preferential alignment with local filament, and whether this preference differs between SRGs and GRGs;
    \item whether filaments imprint morphological asymmetries in SRGs and GRGs, and how these signatures compare across the two populations.
\end{itemize}
All these aspects are very significant to identify any distinctive pattern of interactions of SRGs and GRGs with filament environments, thereby unveiling how cosmic filaments - the principal habitat of GRGs, mediate jet propagation and ultimately regulate their growth and evolution.

The outline of the paper is as follows: In Sec.~\ref{sampelcreation} we describe the construction of  GRG, SRG samples and the filament catalogue.
Sec.~\ref{Analysis} and the subsections therein detail the methods and statistical framework employed for a comprehensive comparative study. This is followed by our results in Sec.~\ref{results}. Sec.~\ref{Discussion} interpret these findings and discusses their implications, and Sec.~\ref{Summary} summarises the main conclusions and outlines future prospects.
 
The flat $\Lambda$ cold dark matter cosmological model has been adopted throughout this paper, based on the Planck results (H$\rm _0$ = 67.8 km s$^{-1}$ Mpc$^{-1}$, $\Omega\rm _m$ = 0.308,  and $\Omega\rm _{\Lambda}$ = 0.692; \citealt{2016A&A...594A..13P}). The images are presented in the J2000 coordinate system. 

\section{Data and sample selection}\label{sampelcreation}
The first subsection is dedicated to detailing the filament catalogue utilised in this study.
The subsequent subsections give a detailed account of the procedures used to construct the GRG and SRG samples.

\subsection{The filament catalogue}
\label{Fil_Samp}

\citet{Tempel2014_fil_cat} constructed a comprehensive catalogue of filamentary structures in the cosmic web by analysing
the spatial distribution of galaxies from the Sloan Digital Sky Survey Data Release 8 (SDSS; \citealt{sdssyork}, \citealt{SDSS_DR8}). The dataset with redshifts accurate to \simi30 \kms, was restricted to galaxies with r-band apparent magnitudes \mr = 17.77 within the redshift range 0.009 $< z \leq$\,0.155. This selection ensured both spectroscopic completeness and a sufficiently dense sampling for filament detection.

The filament identification in \citet{Tempel2014_fil_cat} was carried out using the Bisous model \citep[for details, see][]{Stoica2005,Tempel_Bisous2016}, a statistically rigorous filament-finding algorithm based on marked point processes with interactions. In this approach, the cosmic web is represented as a network of small cylindrical segments that collectively trace elongated overdensities in the galaxy distribution. Each cylinder, with a characteristic radius of $\sim$\,0.5\,$h^{-1}$\,Mpc, acts as a local filament element; their alignment and connectivity are optimised within a Bayesian framework to maximise consistency with the observed galaxy positions. The model employs Markov Chain Monte Carlo sampling to explore filament configurations and generates a three-dimensional probability field describing the filamentary network. From this field, filament spines and local orientation vectors are extracted with sub-megaparsec precision.

Unlike many other filament-finding methods, the Bisous model does not require a smoothed density field, predefined nodes, or assumptions about galaxy luminosity or mass. It works directly with galaxy positions, allowing detection of both strong and weak filaments while remaining robust to survey limitations. Accordingly, the \citet{Tempel2014_fil_cat} catalogue provides a statistically robust and morphologically accurate mapping of the filamentary network, making it particularly well suited for localising RGs within their large-scale environments.

\subsection{The GRG sample}\label{GRG_samp}
To investigate the properties of GRGs within the cosmic web filaments, we compiled a sample of sources lying within the sky coverage of SDSS filaments (Sec.~\ref{Fil_Samp}). For this study, we used the GRG catalogue of \citet{Mostert2024} based on LoTSS DR2 \citep{lotssDR2}. The catalogue contains 11585 GRGs, the majority of which were identified through an automated, multi-stage pipeline combining machine-learning methods, crowd-sourced classifications, and Bayesian statistical modelling, with the remainder drawn from previously published compilations. This modelling accounts for key selection effects, including surface brightness limitations and identification inefficiencies, thereby enabling constraints on the GRG size distribution, comoving number density, and lobe volume-filling fraction within the cosmic web.

\begin{figure*}[htbp]
\centering
\includegraphics[width=0.84\textwidth]{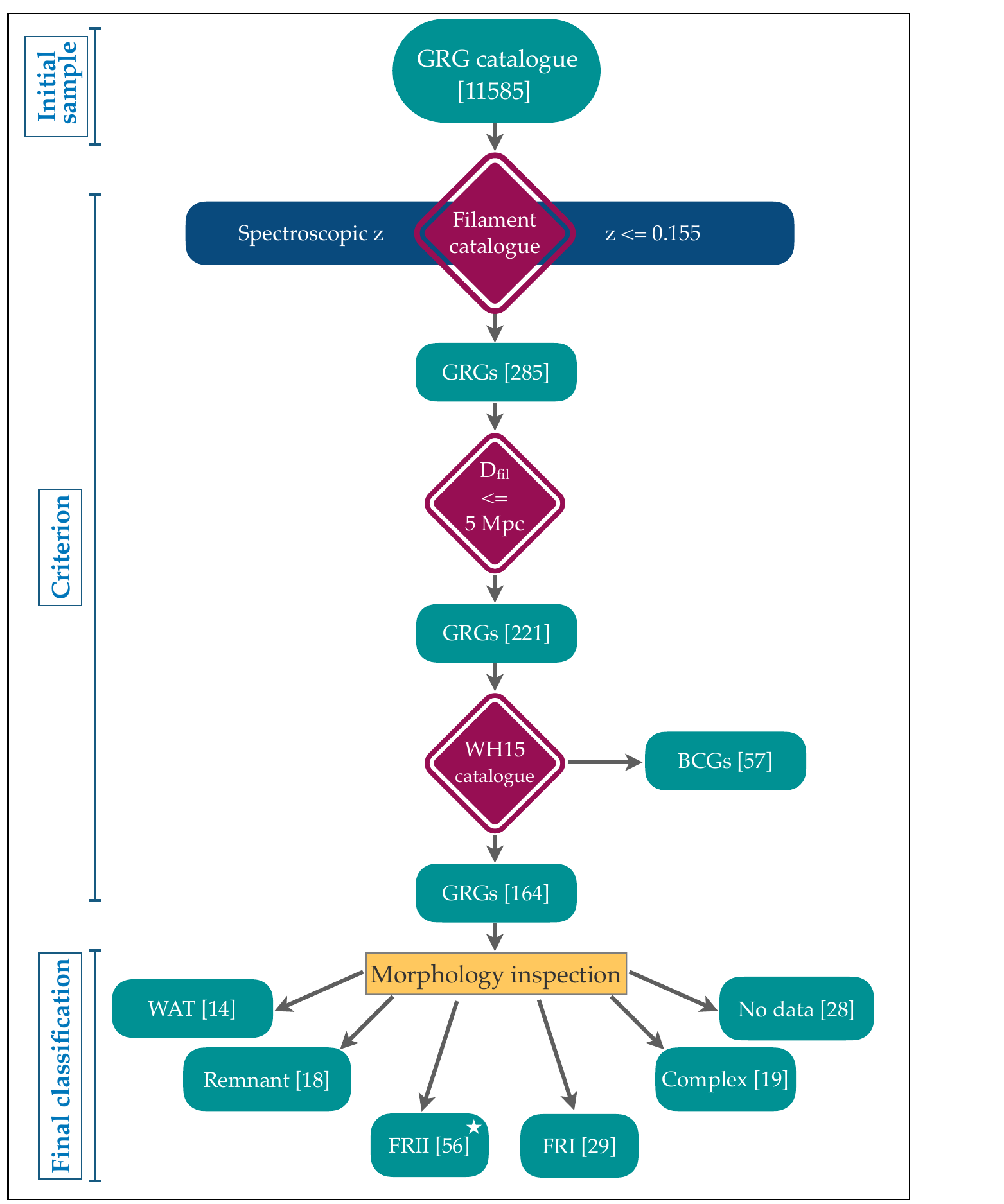}
\caption{The flowchart outlines the construction of the GRG sample, combining the \citet{Mostert2024} GRG catalogue with the filament catalogue of \citet{Tempel2014_fil_cat} and excluding the sources found in the WH15 cluster catalogue. The "No data" category denotes sources lacking publicly available LoTSS data, preventing reliable morphological classification.}
\label{fig:samp_GRG_Fil}
\end{figure*}

From \citet{Mostert2024} catalogue, we extracted a subset of 285 GRGs located within the SDSS filament catalogue footprint \citep{Tempel2014_fil_cat} and have spectroscopic redshifts of $z\leq$\,0.155. Of these, 221 GRGs were found to lie within a 3D comoving distance of 5~Mpc (D$\rm _{fil} \leq$ 5~Mpc) from the cosmic filaments (Fig.~\ref{fig:samp_GRG_Fil}). Given the heterogeneous nature of cosmic filaments, we sought to minimise environmental biases by removing 57 GRGs whose host galaxies are classified as brightest cluster galaxies (BCGs) using WH15 \citep{WH15}. This exclusion ensures that our analysis probes the direct influence of filamentary environments, uncontaminated by the distinct physical conditions of dense cluster cores, and yields a refined working sample of 164 GRGs.

We then performed a detailed morphological classification of each GRG by visually inspecting multi-frequency images from LoTSS DR2 (6$\arcsec$ and 20$\arcsec$ maps at 144 MHz), TGSS ADR1 \citep[$\sim$\,25$\arcsec$ maps at 147 MHz);][]{intema-tgss-17},  FIRST  \citep[5$\arcsec$ maps at 1.4 GHz;][]{beckerfirst95}, NVSS \citep[45$\arcsec$ maps at 1.4 GHz;][]{nvss} and VLASS \citep[2.5$\arcsec$ maps at 3 GHz;][]{vlass}. LoTSS, with its unparalleled surface-brightness sensitivity, reveals the faint diffuse plasma, while VLASS and FIRST, with their higher angular resolution, resolve compact cores, jets, and terminal hotspots. Together, these complementary datasets allow us to trace compact features, bridge or wing emission, and diffuse relic lobes, thereby establishing a near-complete morphological classification for every source.

We then grouped the sources into the following classes, where \textit{N} is the number of sources:

     \begin{itemize}
         \item FRI: centre-brightened jets/plumes ($N = 29$)
         \item FRII: edge-brightened lobes with hotspots ($N = 56$)
         \item WAT: wide-angle tails with C-shaped bends ($N = 14$)
         \item Remnant: lobe-dominated, core-faint systems lacking jets or hotspots ($N = 18$)
         \item Complex: hybrid or ambiguous morphologies that do not fit the above classes ($N = 19$)
     \end{itemize}
Morphological classification was not possible for 28 sources due to the lack of publicly available LoTSS data, labelled as "No data" in the flowchart (Fig.~\ref{fig:samp_GRG_Fil}).
To maintain uniformity and reduce complexity in our subsequent analysis of jet-filament alignment and arm-length ratio, we restrict the study to sources classified as FRII.

\begin{figure}
\includegraphics[width=0.45\textwidth]{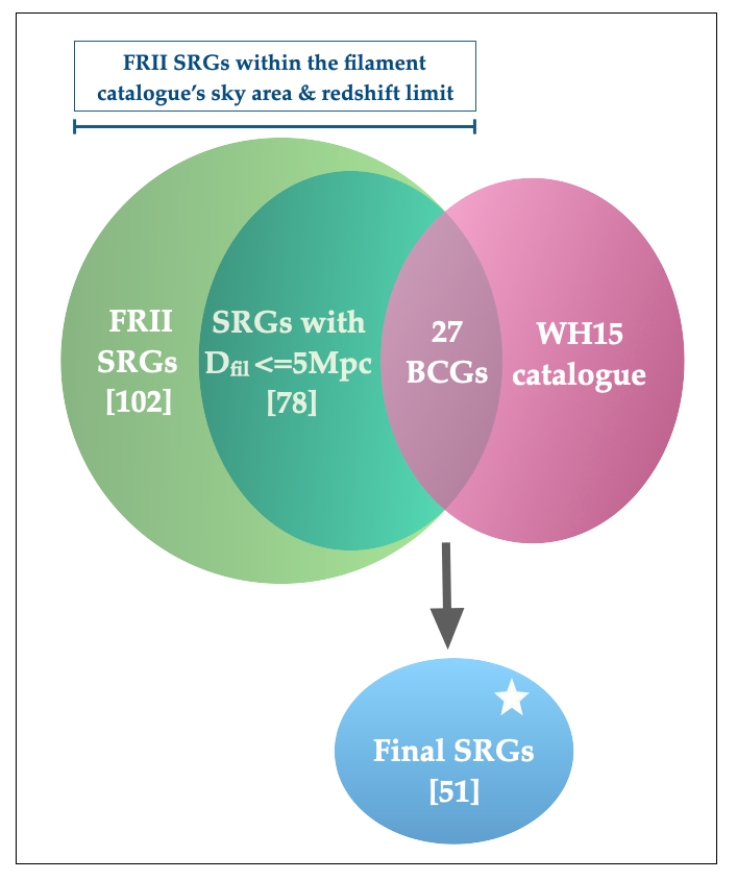}
\caption{The figure illustrates the process used to construct the SRG sample used for our analysis.}
\label{fig:samp_SRG_Fil}
\end{figure} 


\begin{figure}
\includegraphics[width=0.52\textwidth]{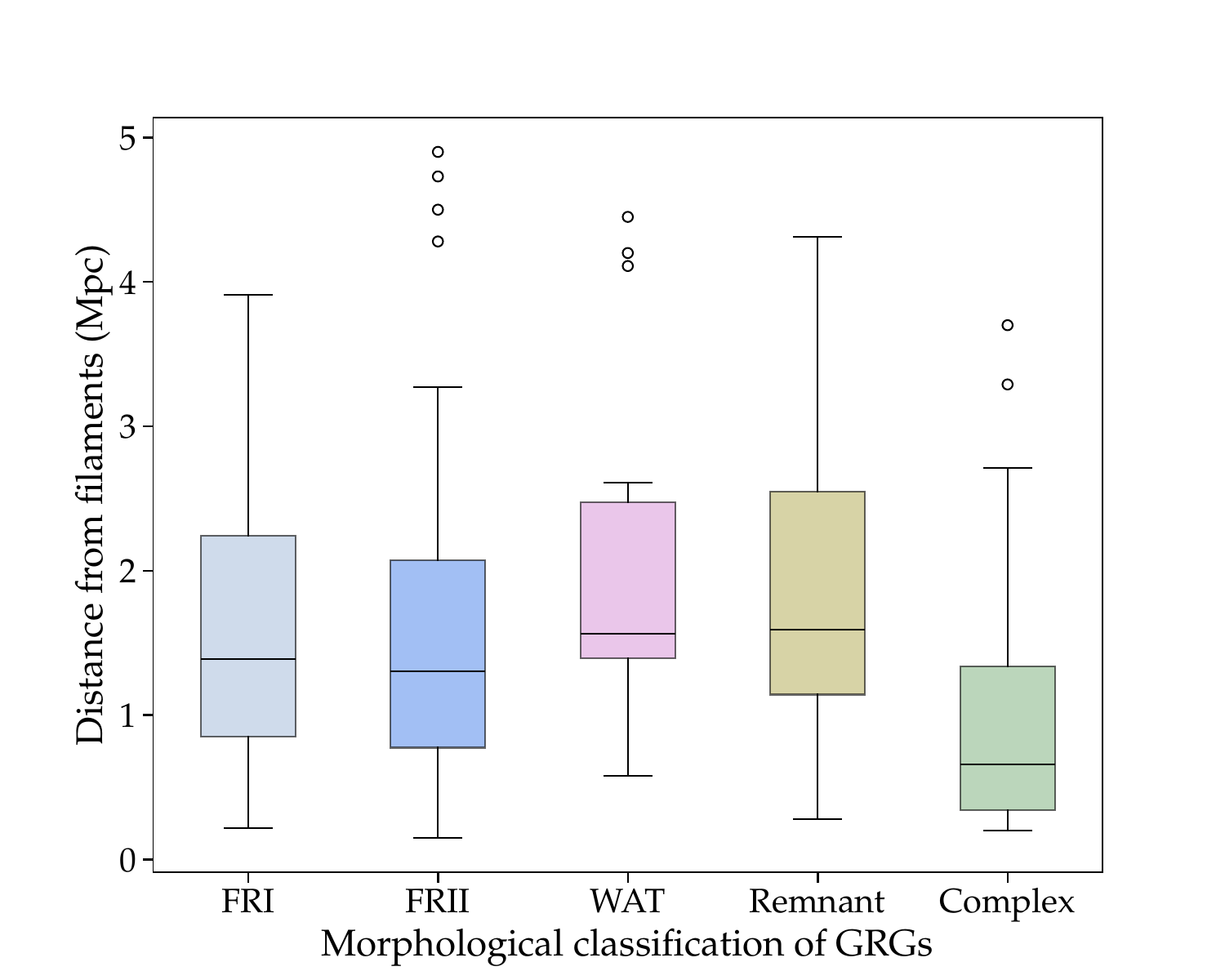}
\caption{The box-and-whisker plots present the distributions of distances from filaments for different morphological classes of GRGs: FRI, FRII, WAT, remnant sources, and complex. The black solid line within a box indicates the median of the distribution, the lower and the upper box boundaries represent the 25th percentile (Q1) and 75th percentile (Q3); therefore, the box shows the interquartile (IQR) range of the data. The whiskers denote the range of the distribution within 1.5 $\times$ IQR, while the circles represent the outliers. As the data are not symmetrically distributed, the whiskers are of unequal length, reflecting the variability above and below the median value. The asymmetry, therefore, indicates skewness or unequal variability in two directions.}
\label{fig:mor_box}
\end{figure} 


\subsection{The SRG sample}
\label{SRG_samp}
To examine whether SRGs exhibit distinct behaviour compared to GRGs within filament environments, we constructed a comparison sample of SRGs based on the catalogue of \citet{Capetti2017}. As with the GRG sample, we restricted the SRGs to those located within the SDSS filament catalogue footprint of \citet{Tempel2014_fil_cat}. This SRG catalogue, derived from NVSS, FIRST and SDSS data, contains 122 FRII-type RGs with $z \leq 0.155$. We visually inspected all FIRST radio maps to confirm the FRII morphology with compact hotspots and to ensure reliable structural identification. Projected linear sizes were then measured from hotspot to hotspot, as appropriate for FRII morphology. To retain consistency with the SRG definition, one object with a size exceeding 0.7 Mpc was excluded. This process yielded a final set of 102 SRGs.

The 3D comoving distances of these SRGs to the nearest filaments from the \citet{Tempel2014_fil_cat} catalogue were then computed. Of 102 sources, 78 lie within 5~Mpc of filaments (D$\rm _{fil}$ $\leq$ 5~Mpc; Fig.~\ref{fig:samp_SRG_Fil}). Cross-matching with the WH15 galaxy cluster catalogue further revealed that 27 of these are hosted by BCGs. Since our aim is to isolate the influence of filaments and avoid biases introduced by dense cluster cores, these BCG-associated SRGs were excluded. The final comparison sample, therefore, consists of 51 FRII SRGs residing in filaments, which we use alongside the GRGs to assess environmental trends.

\subsection{Extent of the filaments}

Unlike virialised haloes, filaments are heterogeneous structures that exhibit a wide range of lengths, widths, and densities. Their lack of spherical symmetry, inherent connectivity, and sensitivity to identification methods pose significant challenges in quantifying how the filamentary environment influences galaxy growth and morphology. The geometry and extent of filaments, particularly their width, set the physical scale over which filaments regulate matter flows, pressure gradients, and gas accretion, thereby impacting the alignment and morphological evolution of galaxies. Filament width refers to the characteristic radial extent from the central spine of a filament to the point where the surrounding matter overdensity, traced by dark matter, gas, or galaxies, drops to a threshold, such as the cosmic mean density \citep{Wang2024} or twice the local background density \citep{Bahe2005}.

The core of a filament is usually considered as the high-density region surrounding the filament spine, where the galaxy and matter overdensity is strongly enhanced compared to the cosmic mean. Observational and simulation studies show that this central region typically extends to a characteristic radius of $\sim$\,1-2 Mpc \citep{Cautun2014, Wang2024}.

Several studies estimate filament widths by stacking filaments in simulations and fitting analytic profiles to their radial density distribution. Using N-body and hydrodynamical simulations, \citet{Colberg2005} and \citet{Dolag2006} fitted isothermal or Navarro-Frenk-White (NFW; \citealt{Navarro1997}) like cylindrical profiles to 3D dark matter and gas density distributions, corresponding to filament diameters of $\sim$\,2-5~Mpc. \citet{Cautun2014} used the NEXUS framework and confirmed these results across different tracers. \citet{Tanimura2020} analysed stacked thermal Sunyaev-Zel’dovich (tSZ) Compton-y and CMB lensing profiles of filaments identified in the SDSS survey. Their modelling with a $\beta$-profile yielded a characteristic full width at half maximum (FWHM) of 2-3 Mpc, providing direct observational evidence for the transverse extent of the dense cores of the cosmic filaments, while the detectable influence of filaments in tSZ and lensing signals extends to about 8 Mpc, reflecting their more diffuse outskirts. \citet{Xia2021}, using the Millennium N-body simulation, showed that the tangential velocity of dark matter around the filaments peaks at radii of the order of 1 to 2~Mpc and declines to negligible levels beyond $\sim$\,2~Mpc, marking the boundary of coherent rotation. Complementing this, \citet{Wang2021} provided observational evidence from SDSS data, detecting a similar rotation curve where galaxy velocities peak around 1 Mpc and fade beyond $\sim$\,2~Mpc. Together, these studies support the interpretation that the core width of cosmic filaments is confined within $\sim$\,1-2~Mpc. With a different approach, \citet{Wang2024} defined the filament radius as the extent at which the logarithmic slope of the radial density profile reaches a minimum, analogous to the splash-back radius in haloes. They constructed stacked galaxy number density profiles around filament spines in cosmological simulations, which yielded a boundary at $\sim$\,0.9-1.3 Mpc, demarcating the transition between the filament spine and its quasi-linear outskirts. This method captures a physically meaningful edge of the gravitational influence of cosmic filaments. 

The study by \citet{Chen2015}, based on the MassiveBlack-II hydrodynamical cosmological simulation \citep{MassiveBlack-II}, demonstrates that the alignment of the galaxy major axes with nearby filament orientations remains statistically significant up to a radial distance of approximately 3.5 Mpc. Within this scale, galaxies, particularly the more massive ones, exhibit a clear tendency to align their major axes along the filament direction, while the alignment signal weakens at larger distances. These findings provide strong evidence that the gravitational and tidal influence of cosmic filaments extends well beyond their dense cores, coherently shaping galaxy orientations 
across several Mpc. Observationally, \citet{Jung2025} combined LoTSS DR2 with SDSS and Dark Energy Spectroscopic Instrument (DESI) Legacy Imaging Surveys \citep{DESI} data, showing that massive galaxies within $\sim$11 Mpc of filaments align their optical major axes with filament orientations. In contrast, AGN jets, which are typically orthogonal to host galaxy major axes, become more randomly oriented within $\sim$8 Mpc of filaments, consistent with chaotic gas accretion and turbulence disrupting stable jet launching.

These studies confirm that, while the dense core of a cosmic filament typically spans $\sim$\,1-2~Mpc, its gravitational tidal field and associated dynamical effects propagate to much larger distances. Signatures such as coherent galaxy spin alignments and anisotropic gas accretion remain detectable several megaparsecs from the filament spine, extending the filament’s influence to $\sim$\,5-8 Mpc. Therefore, in our analysis, we consider GRGs and SRGs within 2~Mpc and 5~Mpc of the filament spine to investigate how the cosmic filament environment impacts their growth and morphology.

\begin{figure}
\includegraphics[scale=0.68]{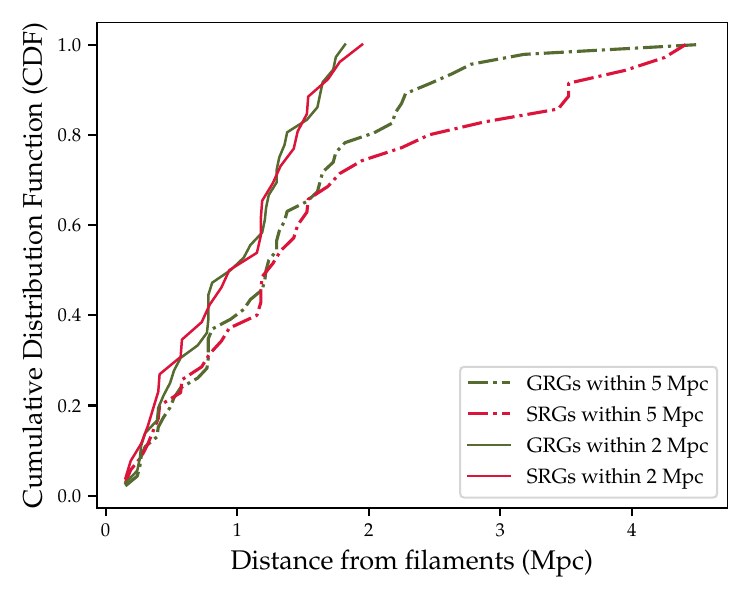}
\caption{The plot shows cumulative distribution functions (CDFs) of the 3D distances to the nearest filament spine for GRGs (green) and SRGs (red). For each population, the solid curves correspond to sources within 2~Mpc of the spine and the dashed curves to sources within 5~Mpc.}
\label{fig:Fil_dist}
\end{figure} 


 \begin{figure*}
\includegraphics[scale=0.45]{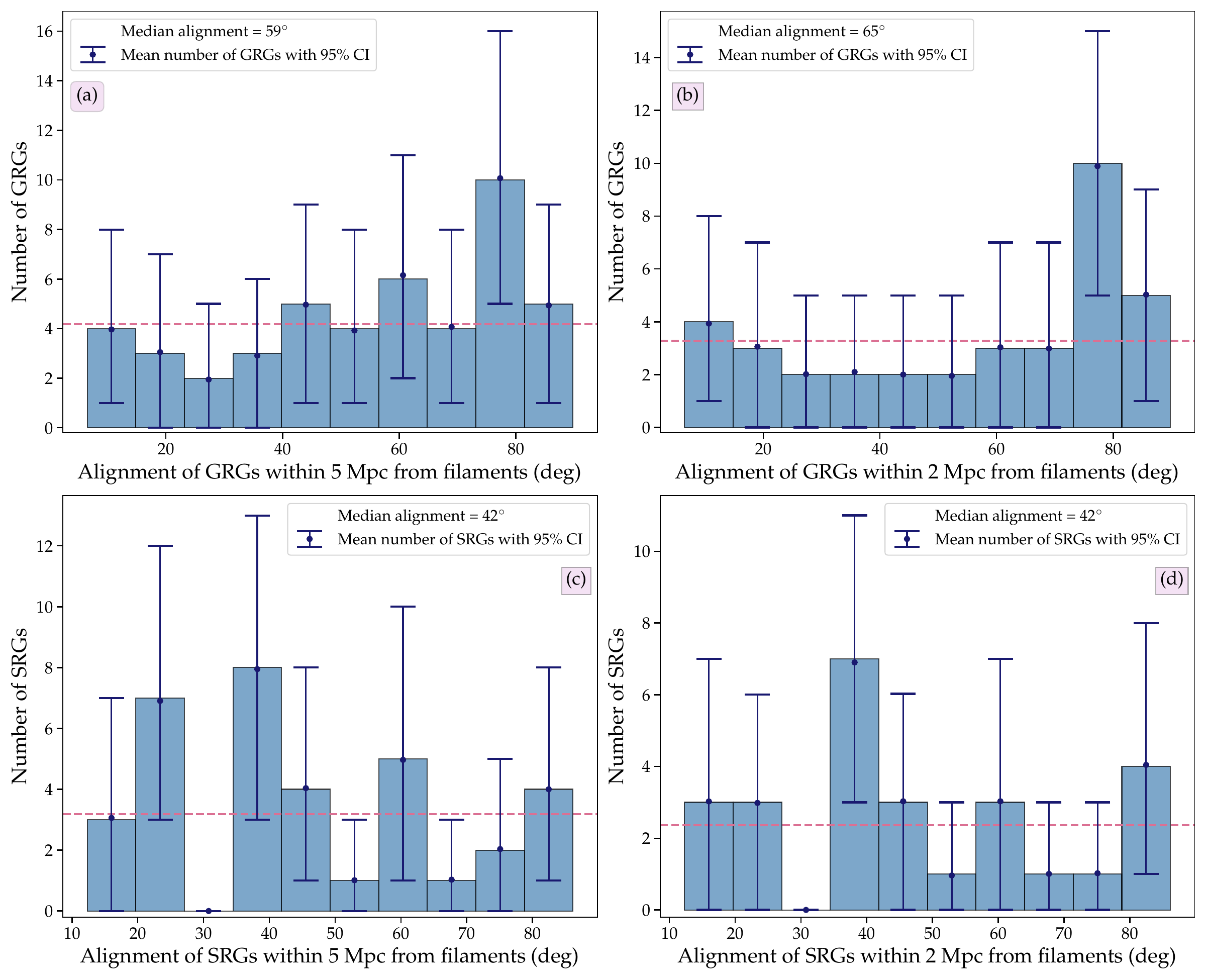}
\caption{The figure shows the distribution of jet-filament alignment angles between the radio jet axis and the local filament spine (0$^{\circ}$ = parallel, 90$^{\circ}$ = perpendicular). Panels (a) and (b) present GRGs within 5 Mpc and 2 Mpc of filaments, with median alignments of 59$^{\circ}$ and 65$^{\circ}$, respectively. Panels (c) and (d) show the corresponding SRG distributions, both with a median of 42$^{\circ}$. The red dashed line shows the distribution for the uniformly oriented samples. The error bars denote 95\% bootstrap (1000 samples) confidence intervals for the mean (blue circles) occupancy of the sources in each bin.}
\label{fig:align_hist}
\end{figure*}


 \begin{figure*}
\includegraphics[scale=0.42]{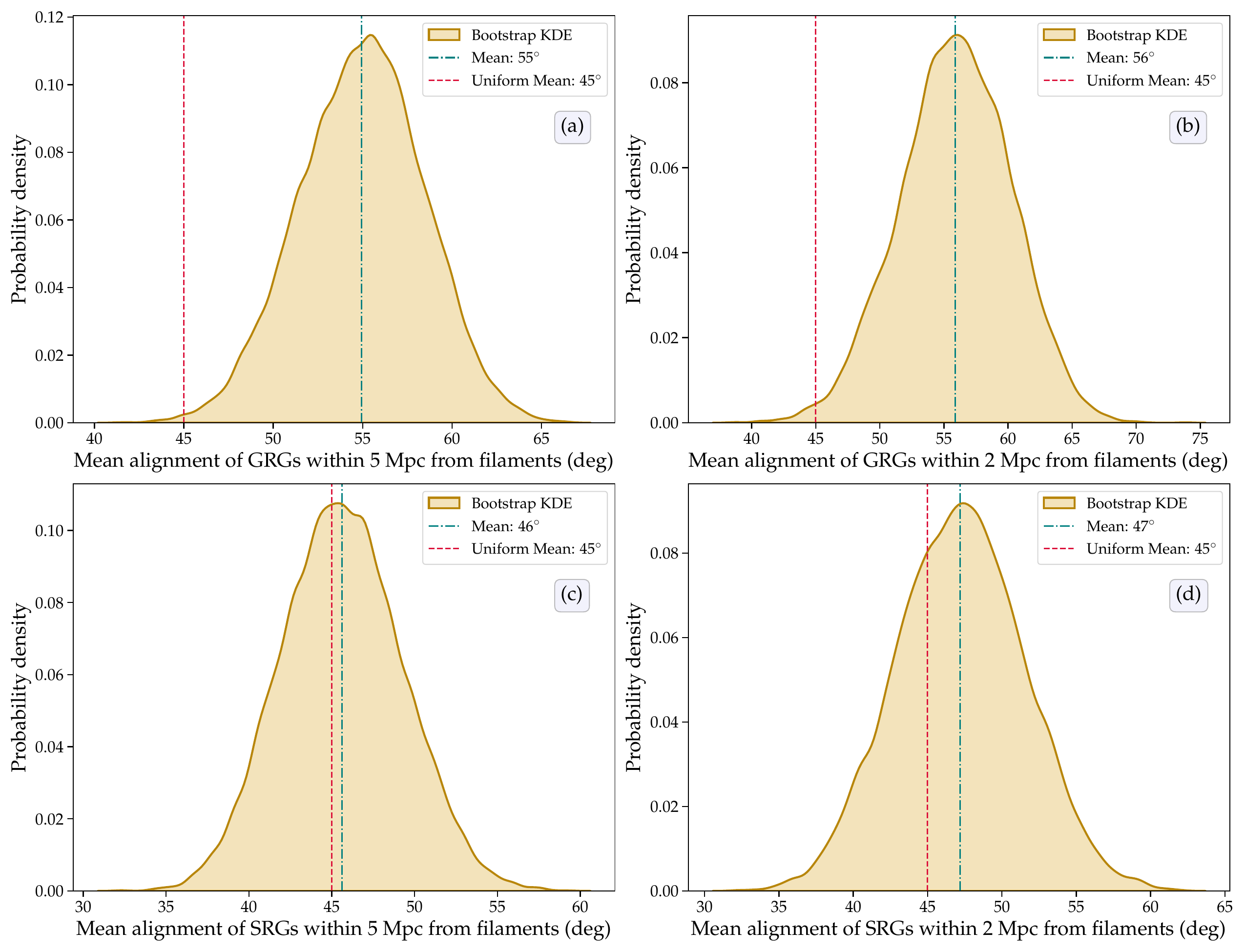}
\caption{The plot shows the kernel density estimations (KDEs) of mean jet-filaments alignments in 1000 bootstrap realisations. Panels (a), (b), (c) and (d) represent the estimates for GRGs within 5~Mpc, GRGs within 2~Mpc, SRGs within 5~Mpc and SRGs within 2~Mpc of filaments. The red dashed vertical line marks the uniform mean, and the green dashed vertical line marks the bootstrap mean from 1000 resamples for the corresponding subsets.}
\label{fig:align_KDE}
\end{figure*}

 \begin{figure}
\includegraphics[scale=0.39]{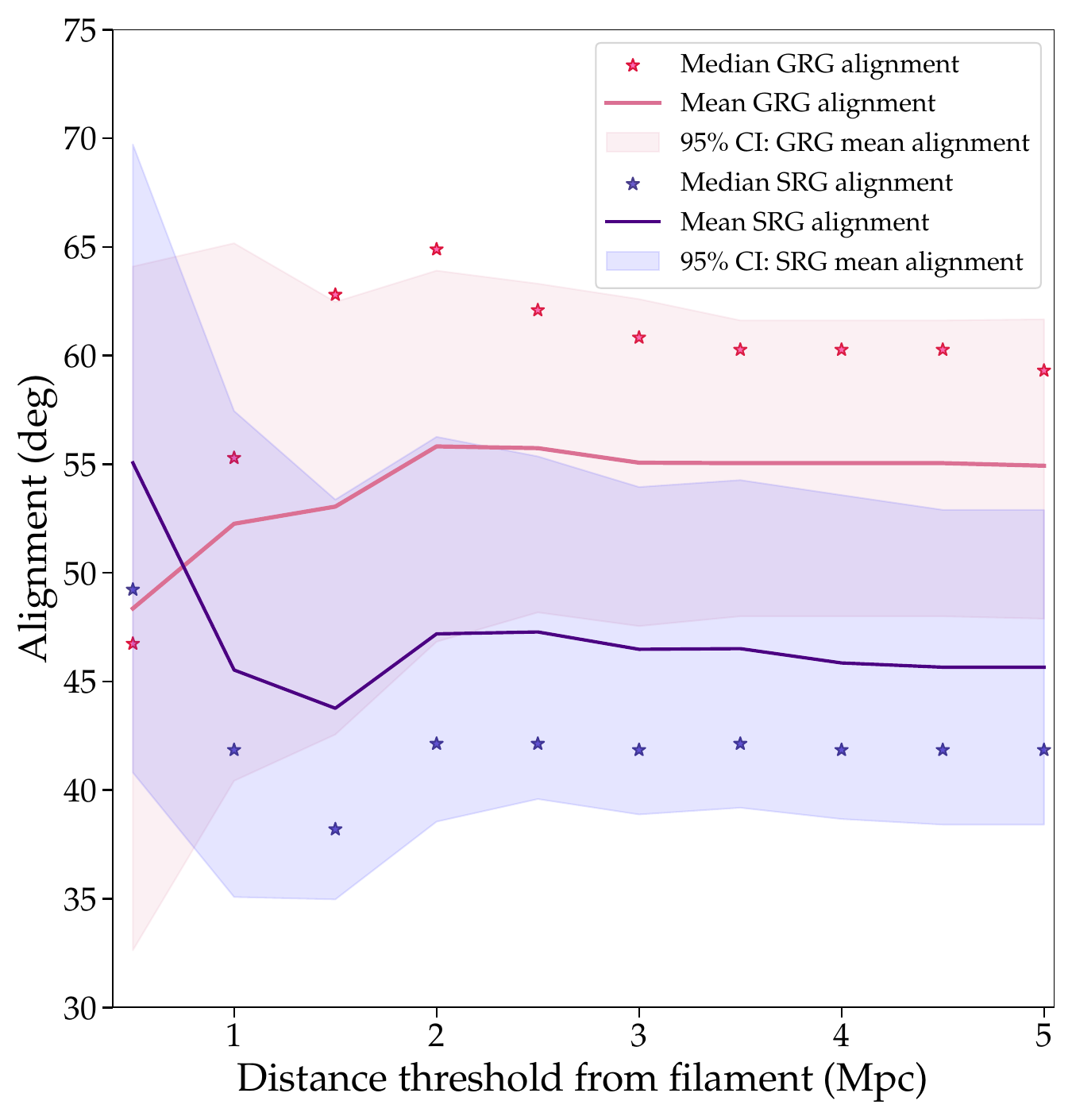}
\caption{The plot shows how GRG and SRG mean jet-alignments change with distance threshold from filaments. The solid red curve traces the GRG mean, the pink shaded region denotes the 95\% bootstrap CI, and red stars mark the median alignment in each distance bin. The solid blue curve represents the SRG mean, the lavender band shows the 95\% bootstrap CI, and blue stars mark the corresponding medians.}
\label{fig:orientation}
\end{figure} 

 \begin{figure}
\includegraphics[scale=0.44]{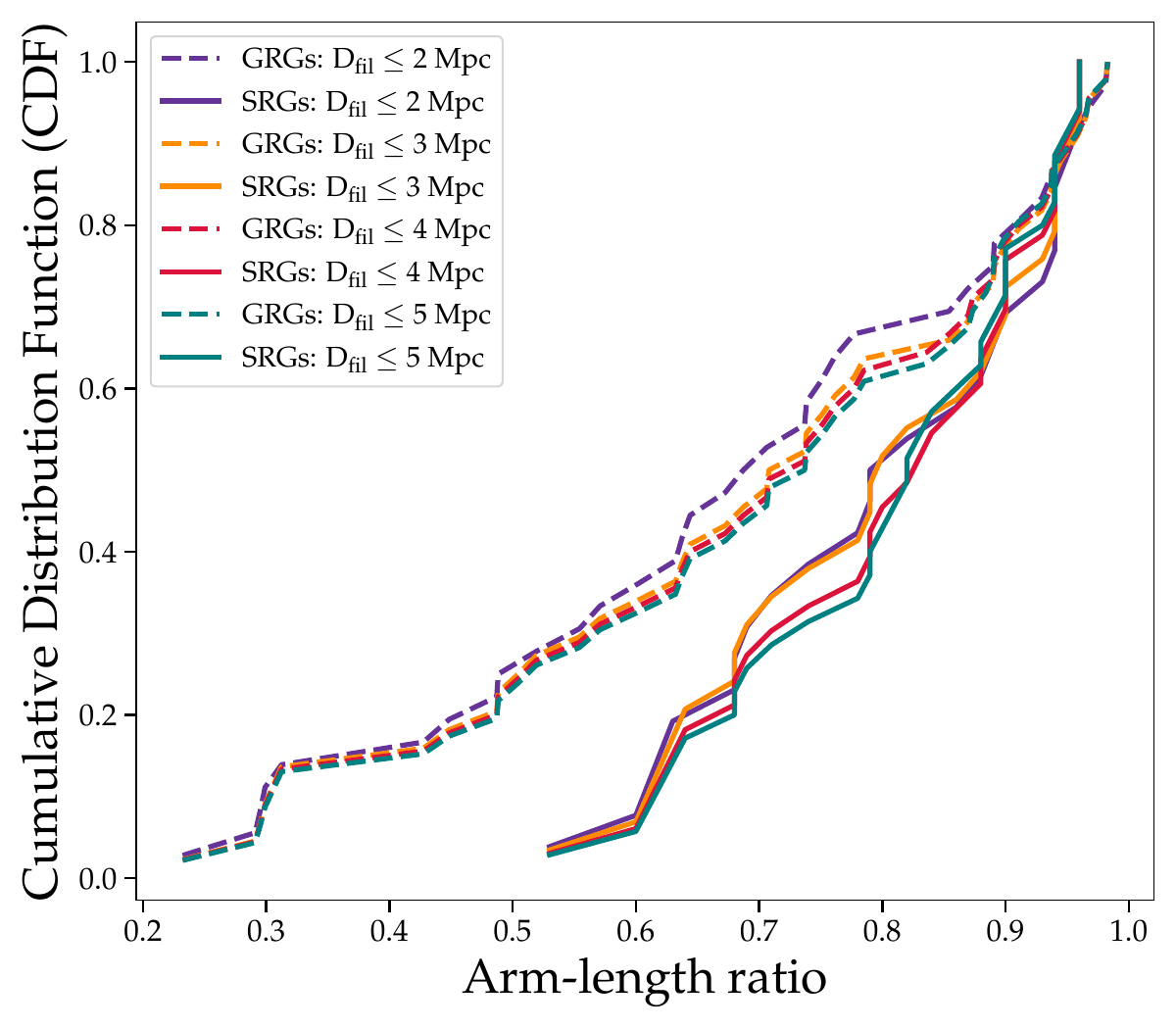}
\caption{The figure displays the cumulative arm-length ratio (ALR) distributions for GRGs (dashed curve), and SRGs (solid curve), stratified by proximity to the nearest filament spine: 2~Mpc (blue), 3 Mpc (orange), 4 Mpc (red), and 5~Mpc (green).}
\label{fig:alr}
\end{figure} 

\section{Analysis of radio galaxies in cosmic filaments}
\label{Analysis}

This section outlines the methodology employed for analysing the properties of GRG and SRG samples in cosmic web filaments.

 \subsection{Distance from the filaments}
 \label{distance}
The distance of a galaxy from the nearest filament serves as a useful statistical proxy, enabling a meaningful investigation into the role of filaments in driving galaxy evolution across the cosmic web. In the \citet{Tempel2014_fil_cat} catalogue, each filament is represented as a spine formed by a sequence of sampling points\footnote{Defined as the set of points used to probe the galaxy distribution for possible filament presence within the Bisous model framework.}. For each of these points, the catalogue provides their 3D coordinates: right ascension (RA), declination (Dec) and redshift ($z$). 
From these data, the Cartesian comoving coordinates ($x_{c}$, $y_{c}$, $z_{c}$) of the sampling points that form the filament spines and the host galaxies of the GRGs were derived using the following:
\begin{align*}
       x_{c} &= D \sin(90^\circ - \delta) \cos\alpha \\
       y_{c} &= D \sin(90^\circ - \delta) \sin\alpha \\
       z_{c} &= D \cos(90^\circ - \delta) 
\end{align*}
 where $\alpha$ $=$ RA, $\delta$ $=$ Dec, and $D$ is the comoving distance. Thereafter, the 3D distances between the GRG host galaxy and filament sampling points were calculated, enabling us to identify the filament closest to the GRG host galaxy by finding the closest sampling point of the filament. Using this approach, we identified the filaments closest to the GRG host galaxies and analysed the distribution of the distances of GRGs from their nearest filaments. A box-and-whisker summary of the 3D distances of different classes of GRGs within 5~Mpc from the nearest filament spine is shown in Fig.~\ref{fig:mor_box}.

For comparison, we also analysed a control sample of SRGs, applying the same methodology to determine their 3D distances from the nearest cosmic filaments. Sources, for which the nearest filament point corresponded to the filament end-point, were excluded from both samples as filament terminations are not well-defined and may introduce ambiguity in assessing the influence of the filamentary environment. In this analysis, we considered FRII GRGs and FRII SRGs located within 5~Mpc (GRGs: 46, SRGs: 35) and 2~Mpc (GRGs: 36, SRGs: 26) of filament spine (see Tab.~\ref{tab:main} in Appendix~\ref{sec:appendix-tab}). The resulting distance distributions are shown in Fig.~\ref{fig:Fil_dist}.

\subsection{Orientation with respect to filaments}
 \label{orientation}

Given the complexity of AGN activity, analysing jet alignments with cosmic filaments in a statistical framework offers a robust means of identifying underlying trends. Such analyses are essential for understanding how anisotropic filamentary environments influence the structural and dynamical evolution of radio AGN.

Since the filaments are inherently curvilinear structures, the filament axis for determining the alignment of radio jets, particularly in FRII GRGs and SRGs, is defined as the projected straight line (on the sky plane) connecting the two sampling points adjacent to the nearest filament sampling point (Sec.~\ref{distance}). Sources for which the nearest filament point lies at the filament terminus were excluded from the alignment analysis, as filament ends are not well defined.

For simplicity, the axis of each GRG (and SRG) is defined as the line segment connecting its two radio hotspots on the sky plane. Therefore, accurately identifying the exact locations of the hotspots is essential for a reliable axis determination. To achieve this, we visually inspected the multi-frequency radio images from multiple radio surveys (see Sec.~\ref{GRG_samp}). The hotspot coordinates (RA, Dec) were taken from the highest-frequency maps available, provided that the compact hotspot emission was detected at a significance level of at least 3$\sigma$ above the rms noise. This has been illustrated in the figures in Sec.~\ref{sec:appendix-im}.

The position angles of the GRGs were derived from the coordinates of their radio hotspots, while the position angles of the filaments were calculated from the 2D (RA, Dec) locations of the filament sampling points that define the local filament axis, as described above. The minimum angle between each GRG axis and its nearest filament was then obtained by taking the absolute difference between these position angles (for radio jet axis position angle uncertainties see Appendix.~\ref {sec:appendix-error}). This angle served as a quantitative measure of the alignment between the radio jet and the filament axis. The same methodology was employed to assess the alignment of SRGs with the filament axis.

As the original sample sizes for both GRGs and SRGs are relatively small, we generated 1000 bootstrap samples for each population to construct empirical sampling distributions, thereby capturing the inherent variability in the data and providing a robust basis for estimating confidence intervals and assessing potential trends. Each bootstrap realisation was constructed by drawing, with replacement, the same number of objects as in the observed sample; we then repeated the identical analysis on every realisation. For each bootstrap realisation, we computed the number of objects in each bin, yielding a distribution of source counts per bin across the 1000 realisations. From these distributions, we obtained, for each bin, the mean bootstrap count and the corresponding 95\% confidence interval. The resulting error bars are generally asymmetric, reflecting the asymmetry of the underlying bootstrap distributions of source counts per bin (Fig.~\ref{fig:align_hist}).

\subsection{Arm-length ratio}
 \label{ALR}

In the simplest scenario, radio galaxies launch bipolar relativistic jets in opposite directions, each transporting comparable amounts of energy, and in principle, producing symmetric radio structures. In practice, however, this symmetry is frequently disrupted by environmental influences. If the jet, or its associated lobe, encounters a denser region of the IGM, it undergoes stronger interactions with the surrounding material, which in turn impedes its propagation. The increased resistance reduces the projected distance that the jet can advance from the radio core, giving rise to a shorter arm. The resulting morphology is thus asymmetric, with one lobe extending farther than the other. Such asymmetry provides a valuable diagnostic for probing the influence of the environment on the properties and evolution of RGs. A widely used parameter to quantify this effect is the arm-length ratio (ALR), defined as the ratio of the projected length of the shorter arm to that of the longer arm.

In our study, we examined the ALRs of FRII-type GRGs and SRGs, excluding sources located near the terminal regions of filaments, to assess the influence of cosmic web filaments on their growth and morphological development. The projected arm lengths were derived from the angular separations of the radio hotspots with respect to the host galaxy. By definition, the arm with the smaller angular separation was taken as the shorter arm, while the larger separation defined the longer arm. The ratio of these two values yields the ALR, which serves as a quantitative measure of morphological asymmetry and enables a direct comparison between GRGs and SRGs.


\section{Results}
\label{results}
\subsection{Distance from the filaments}
\label{results:distance}
The box-and-whisker plot in Fig.~\ref{fig:mor_box} depicts the distribution of comoving 3D distances from filaments for different morphological classes -- FRI, FRII, WAT, remnant, and complex among 136 GRGs located within 5~Mpc of filaments. GRGs with complex morphology exhibit the smallest median distance from filaments (0.66 Mpc), indicating a pronounced tendency for such sources to be located in close proximity to filamentary structures. Such locations, typically associated with enhanced galaxy and gas densities, may facilitate interactions or mergers that could contribute to the development of complex radio structures. In comparison, FRI and FRII sources exhibit median distances of 1.39 Mpc and 1.31 Mpc, respectively; their comparable values suggest similar environmental conditions influencing their growth. WAT and remnant sources lie slightly farther from filaments, with median distances of 1.56 Mpc and 1.59 Mpc, respectively, indicating that these classes may preferentially occur in regions less tightly associated with filamentary spines compared to other morphological types. This environmental preference may reflect different evolutionary stages, with WATs potentially tracing dynamically active group or cluster outskirts \citep{Odea21WAT}, and remnants representing fading RGs \citep{2024Morganti} that have migrated or evolved away from the densest filamentary environments. These systematic differences in median distances across morphological classes highlight the potential role of cosmic web environments in shaping the morphological diversity of GRGs, possibly through a combination of environmental fuelling, dynamical interactions, and evolutionary ageing.

Fig.~\ref{fig:Fil_dist} shows the cumulative distribution functions of distances from cosmic filaments for FRII SRGs and GRGs, considering only sources whose host galaxies are within 2 and 5~Mpc of the filament spines, and excluding sources located at filament ends.  For sources within 2~Mpc of filaments, the median distances are 1.04 Mpc for SRGs and 1.00 Mpc for GRGs; for those within 5~Mpc, the corresponding values are 1.27 Mpc and 1.23 Mpc, respectively. The close agreement between these values, together with the substantial overlap in their cumulative distributions, indicates that SRGs and GRGs exhibit comparable levels of occupancy within the filament environment. This uniformity in large-scale spatial association suggests that proximity to filaments alone is insufficient to account for the size difference between the two populations. Additional factors, such as jet-filament alignment and intrinsic host galaxy properties, are likely to play a more decisive role in determining whether a radio galaxy evolves into a giant.

Additionally, to examine whether the projected linear sizes of RGs depend on their proximity to filaments, we tested for correlations between source size and distance to the filament spine. No statistically significant relationship was found for either the SRG or GRG populations.  Both populations, however, show a mild excess of sources near filament axes, possibly reflecting enhanced fuelling within filament cores that promotes the growth of RGs.

\subsection{Orientation with respect to filaments}
 \label{results:orientation}

The distribution of orientation angles of GRGs, along with their 1000 bootstrap realisations, located within 5~Mpc of filaments is shown in Fig.~\ref{fig:align_hist}a. The median orientation angle of the observational GRG sample (59$^{\circ}$) is in close agreement with that of the corresponding bootstrap median (60$^{\circ}$), indicating a stable alignment at this scale. When the analysis is restricted to GRGs within 2~Mpc of filaments (Fig.~\ref{fig:align_hist}b), the bootstrap median increases slightly to 65$^{\circ}$, remaining consistent with the median value of the observational sample. This similarity across distance thresholds suggests that the observed alignment is robust to variations in the proximity criterion to the filaments. In both figures, the error bars represent the 95\%
confidence interval around the mean number of GRGs per orientation bin, while the red dotted line indicates the expected distribution for a uniform orientation. In both cases, the distributions confirm a tendency for GRGs to be preferentially aligned at larger angles relative to the filament spine.

Fig.~\ref{fig:align_hist}c,d show the orientation distributions of SRGs within 5 and 2~Mpc of filaments, together with 1000 bootstrap samples. In both cases, the median angle is 42$^{\circ}$, identical to the bootstrap median and statistically indistinguishable from the random expectation of 45$^{\circ}$ given the 95\% confidence intervals. Error bars denote the 95\% confidence intervals around the mean counts per bin, and the red dotted line marks the uniform distribution. These results indicate no significant preferential alignment for SRGs, whose orientations remain consistent across distance thresholds. By contrast, GRGs exhibit median offsets by $\sim$15$^{\circ}$ from the random expectation, a deviation significant beyond the 95\% level and consistent with their corresponding mean offsets. The two populations, therefore, show a clear dichotomy: SRGs are consistent with random orientations, while GRGs preferentially align at larger angles to filament spines.

To assess the robustness of the alignment trends, we constructed kernel density estimates (KDEs) of the mean orientations obtained from each bootstrap realisation for both GRGs and SRGs, considering sources located within 5~Mpc and 2~Mpc of the filament spine. For GRGs within 5~Mpc (Fig.~\ref{fig:align_KDE}a), the bootstrap distribution peaks at a mean orientation of $\sim$\,55$^{\circ}$ (green dotted line), which lies well above 45$^{\circ}$, shown by red dotted line - the expectation for a uniform orientation distribution. This offset corresponds to a deviation of $\sim$\,2.9$\sigma$ ($ p\approx$\,0.0042), with 99.8\% of the bootstrap means exceeding 45$^{\circ}$ indicating a statistically significant departure from uniformity. A consistent trend is observed for GRGs within 2~Mpc (Fig.~\ref{fig:align_KDE}b), where the mean orientation is $\sim$\,56$^{\circ}$ with 99.6\% of bootstrap means lying above 45$^{\circ}$, 
and the deviation from uniformity remains highly significant. These results confirm that the preferential alignment of GRGs at larger angles relative to the filament spine is stable across distance thresholds.  
In contrast, SRGs exhibit no such preference. The KDEs of the mean orientations of bootstrapped SRG samples
within 5~Mpc (Fig.~\ref{fig:align_KDE}c) and 2~Mpc (Fig.~\ref{fig:align_KDE}d) peak at $\sim$\,46$^{\circ}$ and $\sim$\,47$^{\circ}$, respectively.  The proportions of bootstrap means above 45$^{\circ}$ are $\sim$\,56\% and $\sim$\,68\%, which are consistent with sampling variability under a uniform distribution. The deviations from uniformity are only  $\sim$\,0.18$\sigma$ ($p\approx$\,0.44), which are statistically insignificant, indicating that SRG jet orientations are consistent with randomness relative to the filament axis.
 
We further examined the mean and median orientations of both GRGs and SRGs as a function of distance from filaments, varying the threshold from 5~Mpc to 0.5 Mpc (Fig.~\ref{fig:orientation}). For thresholds between 2~Mpc and 5~Mpc, GRGs consistently display mean and median orientations significantly above the uniform expectation, while SRGs remain consistent with random alignment within 95\% confidence intervals. Below 2~Mpc, the reduced sample sizes limit any statistically significant inferences.

To quantitatively test the null hypothesis of uniformity, we applied both the Kolmogorov–Smirnov (KS) and Anderson–Darling (AD) tests across the range of distance thresholds. For GRGs, KS test $p$-values satisfy $p \leq 0.02$ and AD test $p \leq 0.01$, providing strong evidence against uniformity. For SRGs, the KS test yields $p \geq 0.4$ and the AD test $p \geq 0.3$, indicating no significant departure from random orientation.

Overall, the bootstrap analyses, distance-dependent trends, and statistical tests consistently reveal a clear dichotomy between the two populations: GRGs show a significant tendency to orient at larger angles relative to the filament spine, whereas SRGs are consistent with random orientation. This contrast likely reflects intrinsic differences in how the two classes interact with the filamentary environment and in their subsequent evolutionary pathways.

Our jet-filament angles are measured in projection on the plane of the sky. For radio galaxies, especially the extended FRII systems that dominate the GRG class, AGN unification \citep[e.g.,][]{AGN_unification}, their low core dominance and well-resolved, bipolar edge-brightened morphologies imply viewing angles close to the sky plane. For such sources, the projected radio axis closely traces the intrinsic jet direction, and hence the projection effect on the jet position angle is minimal. Moreover, the projection of the filament spine removes only the line-of-sight component and therefore tends to move intrinsically extreme alignments towards more moderate apparent angles. Consequently, the large alignment angles we report are conservative lower limits to the actual alignments. We performed bootstrap resampling of the alignment measurements, which yields stable and consistent angle distributions, supporting the robustness of our results.

\subsection{Arm-length ratio}
 \label{results ALR}
Symmetry in RGs is often disrupted by their surrounding environment, making ALR a useful probe of how external conditions influence source morphology. In this section, we examine how ALR varies with filamentary environment, testing whether asymmetries correlate with distance to filament spines. For example, \citet{Pirya2012} reported that in 3C~326 the pronounced arm length asymmetry is linked to a nearby filamentary overdensity, and suggested that similar environmental anisotropies may underlie lobe asymmetries in other RGs.

Fig.~\ref{fig:alr} shows the cumulative distribution functions of ALR for SRGs and GRGs based on their proximity to the filaments. The distribution highlights systematic differences in asymmetry between the two populations across different distance thresholds. For GRGs, the mean and median ALR within 5~Mpc of filaments are 0.69 and 0.73, respectively, decreasing slightly to 0.67 and 0.69 within 2~Mpc. By contrast, SRGs display higher and more consistent values, with a mean and median of 0.81 and 0.82 within 5~Mpc, and 0.80 and 0.80 within 2~Mpc of filaments.
The lower ALR values in GRGs, especially in proximity to filaments, indicate a greater degree of structural asymmetry relative to SRGs. This trend suggests that the larger physical extents of GRGs render them more susceptible to anisotropies in the ambient medium, amplifying the influence of the filamentary environment on their jet propagation and overall morphology. In contrast, the relative stability of ALR in SRGs across filament distances implies a weaker environmental impact.

\section{Discussion: Environmental coupling of radio galaxies within cosmic filaments}
\label{Discussion}

Cosmic web filaments, the principal components of the large-scale structure, trace the distribution of dark and baryonic matter and serve as conduits for gas flows, galaxy evolution, and matter transport. They form a dynamic environment in which galaxies and their radio jets interact with anisotropic gas streams, pressure gradients, and magnetic fields \citep{1996Bondi,2014Dubois,Cautun2014,Wang2024}. Striking individual systems already demonstrate how profoundly filaments and sheets can reshape radio galaxy morphology. The GRG 0503–286, for example, displays a remarkable inversion-symmetric, X-shaped structure, with its lobes sharply bent away from the source axis and separated by a $\sim$\,200 kpc emission-free gap \citep{Dabhade2022XRG}. This unusual morphology cannot be explained by host-centred processes alone; instead, it is best interpreted as the deflection of lobe backflows by a large-scale sheet of the cosmic web, consistent with density asymmetries and Warm Hot Intergalactic Medium (WHIM)-like gas acting across megaparsec scales. Such cases illustrate the rich astrophysics that can be extracted from radio galaxies as probes of the intergalactic medium, revealing the role of sheet- and filament-driven anisotropic pressure, buoyancy, and large-scale tidal fields. Yet, before these environmental imprints can be interpreted at a population level, the essential first step is to robustly identify RGs within filaments and to locate them precisely with respect to the filament spine and boundaries. In this work, we address this gap by identifying RGs within a three-dimensional filamentary framework and systematically examining their properties across distinct filament environments. By doing so, we provide a framework enabling a quantitative assessment of how the cosmic web filament influences their growth, asymmetries, and feedback. 

\subsection{Proximity to filament: impact on morphology}
The analysis of filament environments across different GRG morphological classes reveals that complex sources ($\sim$\,18\% of the sample) exhibit a pronounced preference for close proximity to filament spines, exhibiting a median distance of 0.66 Mpc (Fig.~\ref{fig:mor_box}). At such small separations, the host galaxies reside in regions of elevated matter density and coherent flows characteristic of filament cores. These environments are subject to enhanced gravitational interactions and tidal forces, which can induce asymmetries in the accretion flow, perturb the stability of the radio jets, and distort their propagation paths. Such conditions hinder the development of the symmetric, well-collimated lobes characteristic of classical FR morphologies, instead producing irregular or disrupted radio structures that manifest observationally as complex morphologies. This suggests that the complex class may, at least in part, represent GRGs whose evolutionary trajectory has been shaped by strong environmental modulation within the densest regions of the cosmic web.

The overall filament occupancy of GRGs is not significantly different from that of SRGs (Fig.~\ref{fig:Fil_dist}), indicating that large-scale filament proximity alone does not govern the transition from smaller to giant radio galaxies. Instead, the eventual size and morphology are more likely determined by the interplay of multiple factors, including jet-filament alignment, variations in local galaxy density, episodic AGN activity, and the intrinsic physical properties of the host galaxy.

\subsection{GRG-filament alignment: impact on GRG growth}
Within cosmic web filaments, matter tends to flow along the filament's spine towards denser nodes, such as galaxy clusters and superclusters, where gravitational potential wells are the deepest. Perpendicular to the spine, material from surrounding voids accretes into the filament's high-density axis. These large-scale gravitational flows generate anisotropies in the gas density and pressure around galaxies \citep{Gheller2019}, which can influence AGN fueling by providing a preferred direction for gas inflow, and can also impact jet propagation by creating directional variations in the ambient medium. For RGs residing in such environments, the orientation of the radio axis relative to the filament spine becomes very important for their growth. Jets aligned parallel to the filament may encounter systematically denser, more turbulent media shaped by longitudinal flows toward nodes, whereas jets directed perpendicular to the spine may propagate into comparatively diffuse void-bound regions. Fig.~\ref{fig:Fil_GRG_10}, Fig.~\ref{fig:Fil_GRG_86} and Fig.~\ref{fig:Fil_GRG1} in Appendix.~\ref{sec:appendix-im} are the examples of GRGs aligning nearly parallel, almost perpendicular and at an intermediate angle of 29$^{\circ}$ with their local filament axes. Over the long lifetimes required for GRGs to reach megaparsec scales, these directional differences in the external medium can imprint themselves on source morphology, affecting jet bending, orientation of jet axes, and potentially contributing to variations in ALRs. 

Based on our findings, GRGs tend to exhibit preferentially large angles between their radio jet axes and the filament spine, in contrast to SRGs, which show random orientations. This distinction suggests varied environmental coupling at different physical scales and evolutionary stages of RGs. A GRG oriented at a large angle to the filament spine will likely have one or both jets propagating almost perpendicular to the spine into lower-density, void-facing regions rather than along the denser filament axis as seen in objects studied by \citet{Malarecki2015} and \citet{Oei2024}. Such conditions would reduce resistance to jet advancement, enabling the lobes to expand farther before significant dissipation, thereby facilitating the extreme linear sizes of GRGs. This geometry can facilitate steady lobe advancement over tens to hundreds of millions of years (Myr) required to reach Mpc scales \citep[e.g.,][]{Jamrozy08,Dabhade2023}. 
Such sustained growth also implies a stable jet orientation, which in turn points to prolonged, coherent accretion onto the central engine, the conditions consistent with quasi-steady fuelling modes that preserve jet directionality \citep[e.g.,][]{Liska2019}. Altogether, these results suggest a pathway to the giant regime: RGs with sustained fuelling and jets oriented far from the filament axis are the systems most likely to evolve into GRGs.

\subsection{Environmental gradients: asymmetry in GRGs}
Using our SRG and GRG samples, we tested for statistical trends between jet-filament alignment, ALR and proximity to filament axis, if any, and found no statistically significant correlations in either population. This suggests that the observed lobe asymmetry primarily arises from local conditions such as density variations, small-scale IGM inhomogeneities, and turbulence. Nonetheless, we observe a systematic offset in ALR, with GRGs being more asymmetric on average than SRGs. This is consistent with the fact that GRG lobes traverse longer path lengths through a more anisotropic medium, integrating modest environmental contrasts into appreciable length asymmetries. Consistent with the orientation results, GRGs aligned at large angles to filament spines are likely to encounter steeper transverse density gradients from the spine toward the void-facing regions. Such gradients modulate the advance speeds of the two hotspots differently, enhancing asymmetry in GRGs relative to SRGs, whose jets likely evolve in a more uniform environment with milder pressure differences.

These interpretations are further supported by our ALR statistics.
In a filamentary environment, anisotropic density and pressure fields, in general,  imply that an RG oriented at a large angle to the filament spine will tend to drive its outward lobe into the lower-density inter-filament medium, favouring faster advancement, while its inward lobe encounters the denser spine, resulting in slower advancement and confinement (as seen in asymmetric GRGs shown in Fig.~\ref{fig:Fil_GRG_10} and Fig.~\ref{fig:Fil_GRG_86}). Consistent with this expectation, our GRG sample, already more asymmetric than SRGs, shows 54\% of sources with the shorter arm towards the spine and 46\% with the shorter arm towards the periphery. Sources (e.g., 54\% GRGs) with spine-facing shorter arms experience higher effective ram pressure along their path. Sources (e.g., 46\% GRGs) with a shorter peripheral arm could possibly be explained by local inhomogeneities along the outward path or by interaction with void-to-spine transverse inflow acting as an opposing bulk flow, both consistent with the inhomogeneous nature of filaments. For SRGs, 63\% have the shorter arm away from the spine and 37\% towards the spine. Given their random jet-spine orientations and smaller sizes, the SRG statistics point to local environmental interactions as the primary driver of asymmetry. The absence of pronounced asymmetry in SRGs suggests that filament-driven effects become significant only once jets and lobes extend to Mpc scales; SRGs remain too small to sample these large-scale gradients.

\section{Summary and future prospects}
\label{Summary}

Our analysis focuses on GRGs, selected from \citet{Mostert2024} that lie within cosmic-web filaments traced on the SDSS footprint by \citet{Tempel2014_fil_cat}, and compares them with a control sample of SRGs from \citet{Capetti2017} restricted to the same sky area and environmental selection. After applying the selection criteria (see Fig.~\ref{fig:samp_GRG_Fil}), the final sample comprises 164 GRGs, including 56 of FRII type. Similarly, the \citet{Capetti2017} catalogue provided 51 FRII SRGs used in the analysis.

The results of our study are summarised as follows:
  \begin{itemize}
   \item For both GRGs and SRGs, no statistically significant correlation is found between projected size and 3D comoving distance to the nearest filament spine. However, both populations show a spatial excess within $\sim$2~Mpc of the spine, implying that the enhanced gas supply in filament cores supports the development of extended radio structures.
   
   \item  We find a clear morphological segregation with filament proximity: complex sources occur nearest to the spine, classical FRI/FRII systems at intermediate distances, and WATs and remnants farther out. The denser, more turbulent filament spine region likely promotes anisotropic ram pressure and jet interactions, giving rise to complex structures. At intermediate distances, smoother conditions favour stable jet collimation in FRIIs and gradual deceleration in FRIs. In the outskirts, weaker confinement and crosswinds allow WATs to flare, while the low external pressure enables remnant lobes to expand and persist after jet switch-off.

   \item FRII GRGs exhibit larger alignment angles between their radio jet axes and the filament spine, whereas FRII SRGs display a random orientation distribution. This supports a scenario in which GRG jets preferentially propagate into lower-resistance, void-facing regions, reducing effective ram pressure and facilitating larger growth.

    \item FRII GRGs (lower ALR) are more asymmetric than FRII SRGs, consistent with steeper transverse spine-to-void environmental gradients that differentially modulate hotspots' advance speeds toward and away from the spine, and enhance lobe-length disparities. 

    \item The ALR trends are consistent with the orientation results, indicating that GRGs positioned nearly perpendicular to filament spines experience stronger transverse density gradients, leading to greater lobe asymmetries driven by unequal hotspot advancement.
    
    \end{itemize}

 Our results demonstrate that connecting radio galaxy morphology to the structure of the cosmic web is now within observational reach. By accurately locating GRGs and SRGs within filaments, we establish a framework for probing how the thermodynamic and magnetic properties of large-scale environments shape AGN evolution. This opens the way for a new generation of studies that will move beyond statistical correlations to direct measurements of the intergalactic medium within filaments. Upcoming facilities such as the Square Kilometre Array \citep[SKA\footnote{\url{https://www.skao.int/en}};][]{SKA-Braun19}, LOFAR-2.0\footnote{\url{https://www.lofar.eu/lofar2-0-documentation/}}, and the 4-metre Multi-Object Spectroscopic Telescope (4MOST; \citealt{de_Jong19}), combined with deep X-ray and optical surveys, will enable high-resolution polarimetric and spectral diagnostics capable of tracing pressure, magnetic fields, and feedback processes across the cosmic web. Together, these efforts promise to transform radio galaxies into precision tools for mapping the physical state and evolution of the filamentary Universe.

\begin{acknowledgements}
This work was supported by the Estonian Ministry of Education and Research (grant TK202), Estonian Research Council grants (PRG1006, PRG3034, PSG1045) and the European Union's Horizon Europe research and innovation programme (EXCOSM, grant No. 101159513).
We acknowledge that this work has made use of  \textsc{astropy} \citep{astropy}, \textsc{aplpy} \citep{apl}, \textsc{matplotlib} \citep{plt}, \textsc{topcat} \citep{top05} and the legacy surveys \url{https://www.legacysurvey.org/acknowledgment/}.
\end{acknowledgements}

\bibliographystyle{aa} 
\bibliography{GRG_Fil.bib}

@article{astropy,
	title = "{Astropy: A community Python package for astronomy}",
	author = {{Astropy Collaboration} and {Robitaille}, Thomas P. and {Tollerud}, Erik J. and {Greenfield}, Perry and {Droettboom}, Michael and {Bray}, Erik and {Aldcroft}, Tom and {Davis}, Matt and {Ginsburg}, Adam and {Price-Whelan}, Adrian M. and {Kerzendorf}, Wolfgang E. and {Conley}, Alexander and {Crighton}, Neil and {Barbary}, Kyle and {Muna}, Demitri and {Ferguson}, Henry and {Grollier}, Fr{\'e}d{\'e}ric and {Parikh}, Madhura M. and {Nair}, Prasanth H. and {Unther}, Hans M. and {Deil}, Christoph and {Woillez}, Julien and {Conseil}, Simon and {Kramer}, Roban and {Turner}, James E.~H. and {Singer}, Leo and {Fox}, Ryan and {Weaver}, Benjamin A. and {Zabalza}, Victor and {Edwards}, Zachary I. and {Azalee Bostroem}, K. and {Burke}, D.~J. and {Casey}, Andrew R. and {Crawford}, Steven M. and {Dencheva}, Nadia and {Ely}, Justin and {Jenness}, Tim and {Labrie}, Kathleen and {Lim}, Pey Lian and {Pierfederici}, Francesco and {Pontzen}, Andrew and {Ptak}, Andy and {Refsdal}, Brian and {Servillat}, Mathieu and {Streicher}, Ole},
	year = 2013,
	month = oct,
	journal = {\aap},
	volume = {558},
	pages = {A33},
	doi = {10.1051/0004-6361/201322068},
	eid = {A33},
	archiveprefix = {arXiv},
	eprint = {1307.6212},
	primaryclass = {astro-ph.IM},
	adsurl = {https://ui.adsabs.harvard.edu/abs/2013A&A...558A..33A}
}

@ARTICLE{SDSS_DR8,
       author = {{Aihara}, Hiroaki and {Allende Prieto}, Carlos and {An}, Deokkeun and {Anderson}, Scott F. and {Aubourg}, {\'E}ric and {Balbinot}, Eduardo and {Beers}, Timothy C. and {Berlind}, Andreas A. and {Bickerton}, Steven J. and {Bizyaev}, Dmitry and {Blanton}, Michael R. and {Bochanski}, John J. and {Bolton}, Adam S. and {Bovy}, Jo and {Brandt}, W.~N. and {Brinkmann}, J. and {Brown}, Peter J. and {Brownstein}, Joel R. and {Busca}, Nicolas G. and {Campbell}, Heather and {Carr}, Michael A. and {Chen}, Yanmei and {Chiappini}, Cristina and {Comparat}, Johan and {Connolly}, Natalia and {Cortes}, Marina and {Croft}, Rupert A.~C. and {Cuesta}, Antonio J. and {da Costa}, Luiz N. and {Davenport}, James R.~A. and {Dawson}, Kyle and {Dhital}, Saurav and {Ealet}, Anne and {Ebelke}, Garrett L. and {Edmondson}, Edward M. and {Eisenstein}, Daniel J. and {Escoffier}, Stephanie and {Esposito}, Massimiliano and {Evans}, Michael L. and {Fan}, Xiaohui and {Femen{\'\i}a Castell{\'a}}, Bruno and {Font-Ribera}, Andreu and {Frinchaboy}, Peter M. and {Ge}, Jian and {Gillespie}, Bruce A. and {Gilmore}, G. and {Gonz{\'a}lez Hern{\'a}ndez}, Jonay I. and {Gott}, J. Richard and {Gould}, Andrew and {Grebel}, Eva K. and {Gunn}, James E. and {Hamilton}, Jean-Christophe and {Harding}, Paul and {Harris}, David W. and {Hawley}, Suzanne L. and {Hearty}, Frederick R. and {Ho}, Shirley and {Hogg}, David W. and {Holtzman}, Jon A. and {Honscheid}, Klaus and {Inada}, Naohisa and {Ivans}, Inese I. and {Jiang}, Linhua and {Johnson}, Jennifer A. and {Jordan}, Cathy and {Jordan}, Wendell P. and {Kazin}, Eyal A. and {Kirkby}, David and {Klaene}, Mark A. and {Knapp}, G.~R. and {Kneib}, Jean-Paul and {Kochanek}, C.~S. and {Koesterke}, Lars and {Kollmeier}, Juna A. and {Kron}, Richard G. and {Lampeitl}, Hubert and {Lang}, Dustin and {Le Goff}, Jean-Marc and {Lee}, Young Sun and {Lin}, Yen-Ting and {Long}, Daniel C. and {Loomis}, Craig P. and {Lucatello}, Sara and {Lundgren}, Britt and {Lupton}, Robert H. and {Ma}, Zhibo and {MacDonald}, Nicholas and {Mahadevan}, Suvrath and {Maia}, Marcio A.~G. and {Makler}, Martin and {Malanushenko}, Elena and {Malanushenko}, Viktor and {Mandelbaum}, Rachel and {Maraston}, Claudia and {Margala}, Daniel and {Masters}, Karen L. and {McBride}, Cameron K. and {McGehee}, Peregrine M. and {McGreer}, Ian D. and {M{\'e}nard}, Brice and {Miralda-Escud{\'e}}, Jordi and {Morrison}, Heather L. and {Mullally}, F. and {Muna}, Demitri and {Munn}, Jeffrey A. and {Murayama}, Hitoshi and {Myers}, Adam D. and {Naugle}, Tracy and {Neto}, Angelo Fausti and {Nguyen}, Duy Cuong and {Nichol}, Robert C. and {O'Connell}, Robert W. and {Ogando}, Ricardo L.~C. and {Olmstead}, Matthew D. and {Oravetz}, Daniel J. and {Padmanabhan}, Nikhil and {Palanque-Delabrouille}, Nathalie and {Pan}, Kaike and {Pandey}, Parul and {P{\^a}ris}, Isabelle and {Percival}, Will J. and {Petitjean}, Patrick and {Pfaffenberger}, Robert and {Pforr}, Janine and {Phleps}, Stefanie and {Pichon}, Christophe and {Pieri}, Matthew M. and {Prada}, Francisco and {Price-Whelan}, Adrian M. and {Raddick}, M. Jordan and {Ramos}, Beatriz H.~F. and {Reyl{\'e}}, C{\'e}line and {Rich}, James and {Richards}, Gordon T. and {Rix}, Hans-Walter and {Robin}, Annie C. and {Rocha-Pinto}, Helio J. and {Rockosi}, Constance M. and {Roe}, Natalie A. and {Rollinde}, Emmanuel and {Ross}, Ashley J. and {Ross}, Nicholas P. and {Rossetto}, Bruno M. and {S{\'a}nchez}, Ariel G. and {Sayres}, Conor and {Schlegel}, David J. and {Schlesinger}, Katharine J. and {Schmidt}, Sarah J. and {Schneider}, Donald P. and {Sheldon}, Erin and {Shu}, Yiping and {Simmerer}, Jennifer and {Simmons}, Audrey E. and {Sivarani}, Thirupathi and {Snedden}, Stephanie A. and {Sobeck}, Jennifer S. and {Steinmetz}, Matthias and {Strauss}, Michael A. and {Szalay}, Alexander S. and {Tanaka}, Masayuki and {Thakar}, Aniruddha R. and {Thomas}, Daniel and {Tinker}, Jeremy L. and {Tofflemire}, Benjamin M. and {Tojeiro}, Rita and {Tremonti}, Christy A. and {Vandenberg}, Jan and {Vargas Maga{\~n}a}, M. and {Verde}, Licia and {Vogt}, Nicole P. and {Wake}, David A. and {Wang}, Ji and {Weaver}, Benjamin A. and {Weinberg}, David H. and {White}, Martin and {White}, Simon D.~M. and {Yanny}, Brian and {Yasuda}, Naoki and {Yeche}, Christophe and {Zehavi}, Idit},
        title = "{The Eighth Data Release of the Sloan Digital Sky Survey: First Data from SDSS-III}",
      journal = {\apjs},
         year = 2011,
        month = apr,
       volume = {193},
       number = {2},
          eid = {29},
        pages = {29},
          doi = {10.1088/0067-0049/193/2/29},
archivePrefix = {arXiv},
       eprint = {1101.1559},
 primaryClass = {astro-ph.IM},
       adsurl = {https://ui.adsabs.harvard.edu/abs/2011ApJS..193...29A}
}

@ARTICLE{Stoica2005,
       author = {{Stoica}, R.~S. and {Mart{\'\i}nez}, V.~J. and {Mateu}, J. and {Saar}, E.},
        title = "{Detection of cosmic filaments using the Candy model}",
      journal = {\aap},
         year = 2005,
        month = may,
       volume = {434},
       number = {2},
        pages = {423-432},
          doi = {10.1051/0004-6361:20042409},
archivePrefix = {arXiv},
       eprint = {astro-ph/0405370},
 primaryClass = {astro-ph},
       adsurl = {https://ui.adsabs.harvard.edu/abs/2005A&A...434..423S}
}

@article{beckerfirst95,
	title = "{The FIRST Survey: Faint Images of the Radio Sky at Twenty Centimeters}",
	author = {{Becker}, Robert H. and {White}, Richard L. and {Helfand}, David J.},
	year = 1995,
	month = sep,
	journal = {\apj},
	volume = {450},
	pages = {559},
	doi = {10.1086/176166},
	adsurl = {https://ui.adsabs.harvard.edu/abs/1995ApJ...450..559B}
}

@ARTICLE{Bahe2005,
       author = {{Bahe}, Yannick M. and {Jablonka}, Pascale},
        title = "{Galaxies in the simulated cosmic web: I. Filament identification and their properties}",
      journal = {arXiv e-prints},
         year = 2025,
        month = feb,
          eid = {arXiv:2502.06484},
        pages = {arXiv:2502.06484},
          doi = {10.48550/arXiv.2502.06484},
archivePrefix = {arXiv},
       eprint = {2502.06484},
 primaryClass = {astro-ph.GA},
       adsurl = {https://ui.adsabs.harvard.edu/abs/2025arXiv250206484B}
}

@ARTICLE{Odea21WAT,
       author = {{O'Dea}, Christopher P. and {Baum}, Stefi A.},
        title = "{Wide-Angle-Tail (WAT) Radio Sources}",
      journal = {Galaxies},
         year = 2023,
        month = may,
       volume = {11},
       number = {3},
          eid = {67},
        pages = {67},
          doi = {10.3390/galaxies11030067},
       adsurl = {https://ui.adsabs.harvard.edu/abs/2023Galax..11...67O}
}

@ARTICLE{Dabhade2022XRG,
       author = {{Dabhade}, Pratik and {Krishna}, Gopal},
        title = "{Discovery of X-shaped morphology of the giant radio galaxy 0503-286}",
      journal = {\aap},
         year = 2022,
        month = apr,
       volume = {660},
          eid = {L10},
        pages = {L10},
          doi = {10.1051/0004-6361/202243463},
archivePrefix = {arXiv},
       eprint = {2204.00839},
 primaryClass = {astro-ph.GA},
       adsurl = {https://ui.adsabs.harvard.edu/abs/2022A&A...660L..10D}
}

@ARTICLE{1996Bondi,
       author = {{Bond}, J. Richard and {Kofman}, Lev and {Pogosyan}, Dmitry},
        title = "{How filaments of galaxies are woven into the cosmic web}",
      journal = {\nat},
         year = 1996,
        month = apr,
       volume = {380},
       number = {6575},
        pages = {603-606},
          doi = {10.1038/380603a0},
archivePrefix = {arXiv},
       eprint = {astro-ph/9512141},
 primaryClass = {astro-ph},
       adsurl = {https://ui.adsabs.harvard.edu/abs/1996Natur.380..603B}
}

@ARTICLE{2024Morganti,
       author = {{Morganti}, Raffaella},
        title = "{What Have We Learned about the Life Cycle of Radio Galaxies from New Radio Surveys}",
      journal = {Galaxies},
         year = 2024,
        month = mar,
       volume = {12},
       number = {2},
          eid = {11},
        pages = {11},
          doi = {10.3390/galaxies12020011},
archivePrefix = {arXiv},
       eprint = {2403.13329},
 primaryClass = {astro-ph.GA},
       adsurl = {https://ui.adsabs.harvard.edu/abs/2024Galax..12...11M}
}

@ARTICLE{Blandford2019,
	title = "{Relativistic Jets from Active Galactic Nuclei}",
	author = {{Blandford}, Roger and {Meier}, David and {Readhead}, Anthony},
	year = 2019,
	month = aug,
	journal = {\araa},
	volume = {57},
	pages = {467--509},
	doi = {10.1146/annurev-astro-081817-051948},
	archivePrefix = {arXiv},
	eprint = {1812.06025},
	primaryClass = {astro-ph.HE},
	adsurl = {https://ui.adsabs.harvard.edu/abs/2019ARA&A..57..467B}
}

@ARTICLE{Casadei2024,
       author = {{Casadei}, Samantha and {Capetti}, Alessandro and {Raiteri}, Claudia M. and {Massaro}, Francesco},
        title = "{The large-scale environment of 3CR radio galaxies at z < 0.3}",
      journal = {\aap},
         year = 2024,
        month = apr,
       volume = {684},
          eid = {A159},
        pages = {A159},
          doi = {10.1051/0004-6361/202347525},
archivePrefix = {arXiv},
       eprint = {2402.13859},
 primaryClass = {astro-ph.GA},
       adsurl = {https://ui.adsabs.harvard.edu/abs/2024A&A...684A.159C}
}

@ARTICLE{DESI,
       author = {{Dey}, Arjun and {Schlegel}, David J. and {Lang}, Dustin and {Blum}, Robert and {Burleigh}, Kaylan and {Fan}, Xiaohui and {Findlay}, Joseph R. and {Finkbeiner}, Doug and {Herrera}, David and {Juneau}, St{\'e}phanie and {Landriau}, Martin and {Levi}, Michael and {McGreer}, Ian and {Meisner}, Aaron and {Myers}, Adam D. and {Moustakas}, John and {Nugent}, Peter and {Patej}, Anna and {Schlafly}, Edward F. and {Walker}, Alistair R. and {Valdes}, Francisco and {Weaver}, Benjamin A. and {Y{\`e}che}, Christophe and {Zou}, Hu and {Zhou}, Xu and {Abareshi}, Behzad and {Abbott}, T.~M.~C. and {Abolfathi}, Bela and {Aguilera}, C. and {Alam}, Shadab and {Allen}, Lori and {Alvarez}, A. and {Annis}, James and {Ansarinejad}, Behzad and {Aubert}, Marie and {Beechert}, Jacqueline and {Bell}, Eric F. and {BenZvi}, Segev Y. and {Beutler}, Florian and {Bielby}, Richard M. and {Bolton}, Adam S. and {Brice{\~n}o}, C{\'e}sar and {Buckley-Geer}, Elizabeth J. and {Butler}, Karen and {Calamida}, Annalisa and {Carlberg}, Raymond G. and {Carter}, Paul and {Casas}, Ricard and {Castander}, Francisco J. and {Choi}, Yumi and {Comparat}, Johan and {Cukanovaite}, Elena and {Delubac}, Timoth{\'e}e and {DeVries}, Kaitlin and {Dey}, Sharmila and {Dhungana}, Govinda and {Dickinson}, Mark and {Ding}, Zhejie and {Donaldson}, John B. and {Duan}, Yutong and {Duckworth}, Christopher J. and {Eftekharzadeh}, Sarah and {Eisenstein}, Daniel J. and {Etourneau}, Thomas and {Fagrelius}, Parker A. and {Farihi}, Jay and {Fitzpatrick}, Mike and {Font-Ribera}, Andreu and {Fulmer}, Leah and {G{\"a}nsicke}, Boris T. and {Gaztanaga}, Enrique and {George}, Koshy and {Gerdes}, David W. and {Gontcho}, Satya Gontcho A. and {Gorgoni}, Claudio and {Green}, Gregory and {Guy}, Julien and {Harmer}, Diane and {Hernandez}, M. and {Honscheid}, Klaus and {Huang}, Lijuan Wendy and {James}, David J. and {Jannuzi}, Buell T. and {Jiang}, Linhua and {Joyce}, Richard and {Karcher}, Armin and {Karkar}, Sonia and {Kehoe}, Robert and {Kneib}, Jean-Paul and {Kueter-Young}, Andrea and {Lan}, Ting-Wen and {Lauer}, Tod R. and {Le Guillou}, Laurent and {Le Van Suu}, Auguste and {Lee}, Jae Hyeon and {Lesser}, Michael and {Perreault Levasseur}, Laurence and {Li}, Ting S. and {Mann}, Justin L. and {Marshall}, Robert and {Mart{\'\i}nez-V{\'a}zquez}, C.~E. and {Martini}, Paul and {du Mas des Bourboux}, H{\'e}lion and {McManus}, Sean and {Meier}, Tobias Gabriel and {M{\'e}nard}, Brice and {Metcalfe}, Nigel and {Mu{\~n}oz-Guti{\'e}rrez}, Andrea and {Najita}, Joan and {Napier}, Kevin and {Narayan}, Gautham and {Newman}, Jeffrey A. and {Nie}, Jundan and {Nord}, Brian and {Norman}, Dara J. and {Olsen}, Knut A.~G. and {Paat}, Anthony and {Palanque-Delabrouille}, Nathalie and {Peng}, Xiyan and {Poppett}, Claire L. and {Poremba}, Megan R. and {Prakash}, Abhishek and {Rabinowitz}, David and {Raichoor}, Anand and {Rezaie}, Mehdi and {Robertson}, A.~N. and {Roe}, Natalie A. and {Ross}, Ashley J. and {Ross}, Nicholas P. and {Rudnick}, Gregory and {Safonova}, Sasha and {Saha}, Abhijit and {S{\'a}nchez}, F. Javier and {Savary}, Elodie and {Schweiker}, Heidi and {Scott}, Adam and {Seo}, Hee-Jong and {Shan}, Huanyuan and {Silva}, David R. and {Slepian}, Zachary and {Soto}, Christian and {Sprayberry}, David and {Staten}, Ryan and {Stillman}, Coley M. and {Stupak}, Robert J. and {Summers}, David L. and {Sien Tie}, Suk and {Tirado}, H. and {Vargas-Maga{\~n}a}, Mariana and {Vivas}, A. Katherina and {Wechsler}, Risa H. and {Williams}, Doug and {Yang}, Jinyi and {Yang}, Qian and {Yapici}, Tolga and {Zaritsky}, Dennis and {Zenteno}, A. and {Zhang}, Kai and {Zhang}, Tianmeng and {Zhou}, Rongpu and {Zhou}, Zhimin},
        title = "{Overview of the DESI Legacy Imaging Surveys}",
      journal = {\aj},
         year = 2019,
        month = may,
       volume = {157},
       number = {5},
          eid = {168},
        pages = {168},
          doi = {10.3847/1538-3881/ab089d},
archivePrefix = {arXiv},
       eprint = {1804.08657},
 primaryClass = {astro-ph.IM},
       adsurl = {https://ui.adsabs.harvard.edu/abs/2019AJ....157..168D}
}

@ARTICLE{Capetti2017,
       author = {{Capetti}, A. and {Massaro}, F. and {Baldi}, R.~D.},
        title = "{FRIICAT: A FIRST catalog of FR II radio galaxies}",
      journal = {\aap},
         year = 2017,
        month = may,
       volume = {601},
          eid = {A81},
        pages = {A81},
          doi = {10.1051/0004-6361/201630247},
archivePrefix = {arXiv},
       eprint = {1703.03427},
 primaryClass = {astro-ph.HE},
       adsurl = {https://ui.adsabs.harvard.edu/abs/2017A&A...601A..81C}
}

@ARTICLE{Chen2015,
       author = {{Chen}, Yen-Chi and {Ho}, Shirley and {Tenneti}, Ananth and {Mandelbaum}, Rachel and {Croft}, Rupert and {DiMatteo}, Tiziana and {Freeman}, Peter E. and {Genovese}, Christopher R. and {Wasserman}, Larry},
        title = "{Investigating galaxy-filament alignments in hydrodynamic simulations using density ridges}",
      journal = {\mnras},
         year = 2015,
        month = dec,
       volume = {454},
       number = {3},
        pages = {3341-3350},
          doi = {10.1093/mnras/stv2260},
archivePrefix = {arXiv},
       eprint = {1508.04149},
 primaryClass = {astro-ph.CO},
       adsurl = {https://ui.adsabs.harvard.edu/abs/2015MNRAS.454.3341C}
}

@article{nvss,
	Title = {{The NRAO VLA Sky Survey}},
	Author = {{Condon}, J.~J. and {Cotton}, W.~D. and {Greisen}, E.~W. and {Yin}, Q.~F. and {Perley}, R.~A. and {Taylor}, G.~B. and {Broderick}, J.~J.},
	Year = 1998,
	Month = may,
	Journal = {\aj},
	Volume = 115,
	Pages = {1693--1716},
	Doi = {10.1086/300337},
	Adsurl = {http://adsabs.harvard.edu/abs/1998AJ....115.1693C}
}

@ARTICLE{DabhadeSAGAN20,
	title = "{{\GG{20201013}}Search and analysis of giant radio galaxies with associated nuclei (SAGAN). I. New sample and multi-wavelength studies}",
	author = {{Dabhade}, P. and {Mahato}, M. and {Bagchi}, J. and {Saikia}, D.~J. and {Combes}, F. and {Sankhyayan}, S. and {R{\"o}ttgering}, H.~J.~A. and {Ho}, L.~C. and {Gaikwad}, M. and {Raychaudhury}, S. and {Vaidya}, B. and {Guiderdoni}, B.},
	year = 2020,
	month = oct,
	journal = {\aap},
	volume = {642},
	pages = {A153},
	doi = {10.1051/0004-6361/202038344},
	eid = {A153},
	archivePrefix = {arXiv},
	eprint = {2005.03708},
	primaryClass = {astro-ph.GA},
	adsurl = {https://ui.adsabs.harvard.edu/abs/2020A&A...642A.153D}
}

@article{PDLOTSS,
	title = "{{\GG{20200228}}Giant radio galaxies in the LOFAR Two-metre Sky Survey. I. Radio and environmental properties}",
	author = {{Dabhade}, P. and {R{\"o}ttgering}, H.~J.~A. and {Bagchi}, J. and {Shimwell}, T.~W. and {Hardcastle}, M.~J. and {Sankhyayan}, S. and {Morganti}, R. and {Jamrozy}, M. and {Shulevski}, A. and {Duncan}, K.~J.},
	year = 2020,
	month = mar,
	journal = {\aap},
	volume = {635},
	pages = {A5},
	doi = {10.1051/0004-6361/201935589},
	eid = {A5},
	archiveprefix = {arXiv},
	eprint = {1904.00409},
	primaryclass = {astro-ph.GA},
	adsurl = {https://ui.adsabs.harvard.edu/abs/2020A&A...635A...5D}
}

@ARTICLE{Colberg2005,
       author = {{Colberg}, J{\"o}rg M. and {Krughoff}, K. Simon and {Connolly}, Andrew J.},
        title = "{Intercluster filaments in a {\ensuremath{\Lambda}}CDM Universe}",
      journal = {\mnras},
         year = 2005,
        month = may,
       volume = {359},
       number = {1},
        pages = {272-282},
          doi = {10.1111/j.1365-2966.2005.08897.x},
archivePrefix = {arXiv},
       eprint = {astro-ph/0406665},
 primaryClass = {astro-ph},
       adsurl = {https://ui.adsabs.harvard.edu/abs/2005MNRAS.359..272C}
}

@ARTICLE{Pirya2012,
       author = {{Pirya}, A. and {Saikia}, D.~J. and {Singh}, M. and {Chandola}, H.~C.},
        title = "{A study of the environments of large radio galaxies using SDSS}",
      journal = {\mnras},
         year = 2012,
        month = oct,
       volume = {426},
       number = {1},
        pages = {758-763},
          doi = {10.1111/j.1365-2966.2012.21656.x},
archivePrefix = {arXiv},
       eprint = {1207.1566},
 primaryClass = {astro-ph.CO},
       adsurl = {https://ui.adsabs.harvard.edu/abs/2012MNRAS.426..758P}
}

@ARTICLE{SKA-Braun19,
       author = {{Braun}, Robert and {Bonaldi}, Anna and {Bourke}, Tyler and {Keane}, Evan and {Wagg}, Jeff},
        title = "{Anticipated Performance of the Square Kilometre Array -- Phase 1 (SKA1)}",
      journal = {arXiv e-prints},
         year = 2019,
        month = dec,
          eid = {arXiv:1912.12699},
        pages = {arXiv:1912.12699},
          doi = {10.48550/arXiv.1912.12699},
archivePrefix = {arXiv},
       eprint = {1912.12699},
 primaryClass = {astro-ph.IM},
       adsurl = {https://ui.adsabs.harvard.edu/abs/2019arXiv191212699B}
}

@ARTICLE{2014Dubois,
       author = {{Dubois}, Y. and {Pichon}, C. and {Welker}, C. and {Le Borgne}, D. and {Devriendt}, J. and {Laigle}, C. and {Codis}, S. and {Pogosyan}, D. and {Arnouts}, S. and {Benabed}, K. and {Bertin}, E. and {Blaizot}, J. and {Bouchet}, F. and {Cardoso}, J. -F. and {Colombi}, S. and {de Lapparent}, V. and {Desjacques}, V. and {Gavazzi}, R. and {Kassin}, S. and {Kimm}, T. and {McCracken}, H. and {Milliard}, B. and {Peirani}, S. and {Prunet}, S. and {Rouberol}, S. and {Silk}, J. and {Slyz}, A. and {Sousbie}, T. and {Teyssier}, R. and {Tresse}, L. and {Treyer}, M. and {Vibert}, D. and {Volonteri}, M.},
        title = "{Dancing in the dark: galactic properties trace spin swings along the cosmic web}",
      journal = {\mnras},
         year = 2014,
        month = oct,
       volume = {444},
       number = {2},
        pages = {1453-1468},
          doi = {10.1093/mnras/stu1227},
archivePrefix = {arXiv},
       eprint = {1402.1165},
 primaryClass = {astro-ph.CO},
       adsurl = {https://ui.adsabs.harvard.edu/abs/2014MNRAS.444.1453D}
}

@ARTICLE{Cautun2014,
       author = {{Cautun}, Marius and {van de Weygaert}, Rien and {Jones}, Bernard J.~T. and {Frenk}, Carlos S.},
        title = "{Evolution of the cosmic web}",
      journal = {\mnras},
         year = 2014,
        month = jul,
       volume = {441},
       number = {4},
        pages = {2923-2973},
          doi = {10.1093/mnras/stu768},
archivePrefix = {arXiv},
       eprint = {1401.7866},
 primaryClass = {astro-ph.CO},
       adsurl = {https://ui.adsabs.harvard.edu/abs/2014MNRAS.441.2923C}
}

@ARTICLE{Jung2025,
       author = {{Jung}, S. Lyla and {Whittam}, I.~H. and {Jarvis}, M.~J. and {Hale}, C.~L. and {Tudorache}, M.~N. and {Yasin}, T.},
        title = "{On the relationship between the cosmic web and the alignment of galaxies and AGN jets}",
      journal = {\mnras},
         year = 2025,
        month = may,
       volume = {539},
       number = {3},
        pages = {2362-2379},
          doi = {10.1093/mnras/staf613},
archivePrefix = {arXiv},
       eprint = {2502.03730},
 primaryClass = {astro-ph.GA},
       adsurl = {https://ui.adsabs.harvard.edu/abs/2025MNRAS.539.2362J}
}

@ARTICLE{Dolag2006,
       author = {{Dolag}, K. and {Meneghetti}, M. and {Moscardini}, L. and {Rasia}, E. and {Bonaldi}, A.},
        title = "{Simulating the physical properties of dark matter and gas inside the cosmic web}",
      journal = {\mnras},
         year = 2006,
        month = aug,
       volume = {370},
       number = {2},
        pages = {656-672},
          doi = {10.1111/j.1365-2966.2006.10511.x},
archivePrefix = {arXiv},
       eprint = {astro-ph/0511357},
 primaryClass = {astro-ph},
       adsurl = {https://ui.adsabs.harvard.edu/abs/2006MNRAS.370..656D}
}

@ARTICLE{sagan5,
       author = {{Dabhade}, P. and {Chavan}, K. and {Saikia}, D.~J. and {Oei}, M.~S.~S.~L. and {R{\"o}ttgering}, H.~J.~A.},
        title = "{Search and analysis of giant radio galaxies with associated nuclei (SAGAN): V. Study of giant double-double radio galaxies from LoTSS DR2}",
      journal = {\aap},
         year = 2025,
        month = apr,
       volume = {696},
          eid = {A97},
        pages = {A97},
          doi = {10.1051/0004-6361/202451918},
archivePrefix = {arXiv},
       eprint = {2408.13607},
 primaryClass = {astro-ph.GA},
       adsurl = {https://ui.adsabs.harvard.edu/abs/2025A&A...696A..97D}
}

@ARTICLE{Dabhade2023,
       author = {{Dabhade}, Pratik and {Saikia}, D.~J. and {Mahato}, Mousumi},
        title = "{Decoding the giant extragalactic radio sources}",
      journal = {Journal of Astrophysics and Astronomy},
         year = 2023,
        month = jun,
       volume = {44},
       number = {1},
          eid = {13},
        pages = {13},
          doi = {10.1007/s12036-022-09898-5},
archivePrefix = {arXiv},
       eprint = {2208.02130},
 primaryClass = {astro-ph.GA},
       adsurl = {https://ui.adsabs.harvard.edu/abs/2023JApA...44...13D}
}

@article{Oei2024,
  title   = {Black hole jets on the scale of the cosmic web},
  author  = {Oei, Martijn S. S. L. and Hardcastle, Martin J. and Timmerman, Roland and others},
  journal = {Nature},
  volume  = {633},
  pages   = {537--541},
  year    = {2024},
  doi     = {10.1038/s41586-024-07879-y}
}

@ARTICLE{FR74,
	title = "{The morphology of extragalactic radio sources of high and low luminosity}",
	author = {{Fanaroff}, B.~L. and {Riley}, J.~M.},
	year = 1974,
	month = may,
	journal = {\mnras},
	volume = {167},
	pages = {31P-36P},
	doi = {10.1093/mnras/167.1.31P},
	adsurl = {https://ui.adsabs.harvard.edu/abs/1974MNRAS.167P..31F}
}

@ARTICLE{Gopal1989,
	title = "{The formation, numbers and radio output of giant radio galaxies}",
	author = {{Gopal-Krishna} and {Wiita}, Paul J. and {Saripalli}, L.},
	year = 1989,
	month = jul,
	journal = {\mnras},
	volume = {239},
	pages = {173--187},
	doi = {10.1093/mnras/239.1.173},
	adsurl = {https://ui.adsabs.harvard.edu/abs/1989MNRAS.239..173G}
}

@ARTICLE{Gheller2019,
       author = {{Gheller}, C. and {Vazza}, F.},
        title = "{A survey of the thermal and non-thermal properties of cosmic filaments}",
      journal = {\mnras},
         year = 2019,
        month = jun,
       volume = {486},
       number = {1},
        pages = {981-1002},
          doi = {10.1093/mnras/stz843},
archivePrefix = {arXiv},
       eprint = {1903.08401},
 primaryClass = {astro-ph.CO},
       adsurl = {https://ui.adsabs.harvard.edu/abs/2019MNRAS.486..981G}
}

@ARTICLE{Liska2019,
       author = {{Liska}, M. and {Tchekhovskoy}, A. and {Ingram}, A. and {van der Klis}, M.},
        title = "{Bardeen-Petterson alignment, jets, and magnetic truncation in GRMHD simulations of tilted thin accretion discs}",
      journal = {\mnras},
         year = 2019,
        month = jul,
       volume = {487},
       number = {1},
        pages = {550-561},
          doi = {10.1093/mnras/stz834},
archivePrefix = {arXiv},
       eprint = {1810.00883},
 primaryClass = {astro-ph.HE},
       adsurl = {https://ui.adsabs.harvard.edu/abs/2019MNRAS.487..550L}
}

@article{plt,
	title = "{Matplotlib: A 2D Graphics Environment}",
	author = {{Hunter}, John D.},
	year = 2007,
	month = may,
	journal = {Computing in Science and Engineering},
	volume = {9},
	number = {3},
	pages = {90--95},
	doi = {10.1109/MCSE.2007.55},
	adsurl = {https://ui.adsabs.harvard.edu/abs/2007CSE.....9...90H}
}

@ARTICLE{intema-tgss-17,
	title = "{The GMRT 150 MHz all-sky radio survey. First alternative data release TGSS ADR1}",
	author = {{Intema}, H.~T. and {Jagannathan}, P. and {Mooley}, K.~P. and {Frail}, D.~A.},
	year = 2017,
	month = feb,
	journal = {\aap},
	volume = {598},
	pages = {A78},
	doi = {10.1051/0004-6361/201628536},
	eid = {A78},
	archivePrefix = {arXiv},
	eprint = {1603.04368},
	primaryClass = {astro-ph.CO},
	adsurl = {https://ui.adsabs.harvard.edu/abs/2017A&A...598A..78I}
}

@ARTICLE{Jamrozy08,
	title = {{A multifrequency study of giant radio sources - II. Spectral ageing analysis of the lobes of selected sources}},
	author = {{Jamrozy}, M. and {Konar}, C. and {Machalski}, J. and {Saikia}, D.~J.},
	year = 2008,
	month = apr,
	journal = {\mnras},
	volume = 385,
	pages = {1286--1296},
	doi = {10.1111/j.1365-2966.2007.12772.x},
	adsurl = {http://adsabs.harvard.edu/abs/2008MNRAS.385.1286J},
	archiveprefix = {arXiv},
	eprint = {0712.0162}
}

@ARTICLE{vlass,
       author = {{Lacy}, M. and {Baum}, S.~A. and {Chandler}, C.~J. and {Chatterjee}, S. and {Clarke}, T.~E. and {Deustua}, S. and {English}, J. and {Farnes}, J. and {Gaensler}, B.~M. and {Gugliucci}, N. and {Hallinan}, G. and {Kent}, B.~R. and {Kimball}, A. and {Law}, C.~J. and {Lazio}, T.~J.~W. and {Marvil}, J. and {Mao}, S.~A. and {Medlin}, D. and {Mooley}, K. and {Murphy}, E.~J. and {Myers}, S. and {Osten}, R. and {Richards}, G.~T. and {Rosolowsky}, E. and {Rudnick}, L. and {Schinzel}, F. and {Sivakoff}, G.~R. and {Sjouwerman}, L.~O. and {Taylor}, R. and {White}, R.~L. and {Wrobel}, J. and {Andernach}, H. and {Beasley}, A.~J. and {Berger}, E. and {Bhatnager}, S. and {Birkinshaw}, M. and {Bower}, G.~C. and {Brandt}, W.~N. and {Brown}, S. and {Burke-Spolaor}, S. and {Butler}, B.~J. and {Comerford}, J. and {Demorest}, P.~B. and {Fu}, H. and {Giacintucci}, S. and {Golap}, K. and {G{\"u}th}, T. and {Hales}, C.~A. and {Hiriart}, R. and {Hodge}, J. and {Horesh}, A. and {Ivezi{\'c}}, {\v{Z}}. and {Jarvis}, M.~J. and {Kamble}, A. and {Kassim}, N. and {Liu}, X. and {Loinard}, L. and {Lyons}, D.~K. and {Masters}, J. and {Mezcua}, M. and {Moellenbrock}, G.~A. and {Mroczkowski}, T. and {Nyland}, K. and {O'Dea}, C.~P. and {O'Sullivan}, S.~P. and {Peters}, W.~M. and {Radford}, K. and {Rao}, U. and {Robnett}, J. and {Salcido}, J. and {Shen}, Y. and {Sobotka}, A. and {Witz}, S. and {Vaccari}, M. and {van Weeren}, R.~J. and {Vargas}, A. and {Williams}, P.~K.~G. and {Yoon}, I.},
        title = "{The Karl G. Jansky Very Large Array Sky Survey (VLASS). Science Case and Survey Design}",
      journal = {\pasp},
         year = 2020,
        month = mar,
       volume = {132},
       number = {1009},
          eid = {035001},
        pages = {035001},
          doi = {10.1088/1538-3873/ab63eb},
archivePrefix = {arXiv},
       eprint = {1907.01981},
 primaryClass = {astro-ph.IM},
       adsurl = {https://ui.adsabs.harvard.edu/abs/2020PASP..132c5001L}
}

@ARTICLE{Lan2021,
       author = {{Lan}, Ting-Wen and {Prochaska}, J. Xavier},
        title = "{On the environments of giant radio galaxies}",
      journal = {\mnras},
         year = 2021,
        month = apr,
       volume = {502},
       number = {4},
        pages = {5104-5114},
          doi = {10.1093/mnras/stab297},
archivePrefix = {arXiv},
       eprint = {2009.04482},
 primaryClass = {astro-ph.GA},
       adsurl = {https://ui.adsabs.harvard.edu/abs/2021MNRAS.502.5104L}
}

@ARTICLE{mack98,
       author = {{Mack}, K. -H. and {Klein}, U. and {O'Dea}, C.~P. and {Willis}, A.~G. and {Saripalli}, L.},
        title = "{Spectral indices, particle ages, and the ambient medium of giant radio galaxies}",
      journal = {\aap},
         year = 1998,
        month = jan,
       volume = {329},
        pages = {431-442},
       adsurl = {https://ui.adsabs.harvard.edu/abs/1998A&A...329..431M}
}

@ARTICLE{Malarecki2015,
       author = {{Malarecki}, J.~M. and {Jones}, D.~H. and {Saripalli}, L. and {Staveley-Smith}, L. and {Subrahmanyan}, R.},
        title = "{Giant radio galaxies - II. Tracers of large-scale structure}",
      journal = {\mnras},
         year = 2015,
        month = may,
       volume = {449},
       number = {1},
        pages = {955-986},
          doi = {10.1093/mnras/stv273},
archivePrefix = {arXiv},
       eprint = {1502.03954},
 primaryClass = {astro-ph.GA},
       adsurl = {https://ui.adsabs.harvard.edu/abs/2015MNRAS.449..955M}
}

@ARTICLE{Massaro2019,
       author = {{Massaro}, F. and {{\'A}lvarez-Crespo}, N. and {Capetti}, A. and {Baldi}, R.~D. and {Pillitteri}, I. and {Campana}, R. and {Paggi}, A.},
        title = "{Deciphering the Large-scale Environment of Radio Galaxies in the Local Universe: Where Are They Born? Where Do They Grow? Where Do They Die?}",
      journal = {\apjs},
         year = 2019,
        month = feb,
       volume = {240},
       number = {2},
          eid = {20},
        pages = {20},
          doi = {10.3847/1538-4365/aaf1c7},
archivePrefix = {arXiv},
       eprint = {1811.11179},
 primaryClass = {astro-ph.HE},
       adsurl = {https://ui.adsabs.harvard.edu/abs/2019ApJS..240...20M}
}

@ARTICLE{Oei2024env,
       author = {{Oei}, Martijn S.~S.~L. and {van Weeren}, Reinout J. and {Hardcastle}, Martin J. and {Gast}, Aivin R.~D.~J.~G.~I.~B. and {Leclercq}, Florent and {R{\"o}ttgering}, Huub J.~A. and {Dabhade}, Pratik and {Shimwell}, Tim W. and {Botteon}, Andrea},
        title = "{Luminous giants populate the dense Cosmic Web. The radio luminosity-environmental density relation for radio galaxies in action}",
      journal = {\aap},
         year = 2024,
        month = jun,
       volume = {686},
          eid = {A137},
        pages = {A137},
          doi = {10.1051/0004-6361/202347115},
archivePrefix = {arXiv},
       eprint = {2404.17776},
 primaryClass = {astro-ph.CO},
       adsurl = {https://ui.adsabs.harvard.edu/abs/2024A&A...686A.137O}
}

@ARTICLE{Mostert2024,
       author = {{Mostert}, R.~I.~J. and {Oei}, M.~S.~S.~L. and {Barkus}, B. and {Alegre}, L. and {Hardcastle}, M.~J. and {Duncan}, K.~J. and {R{\"o}ttgering}, H.~J.~A. and {van Weeren}, R.~J. and {Horton}, M.},
        title = "{Constraining the giant radio galaxy population with machine learning and Bayesian inference}",
      journal = {\aap},
         year = 2024,
        month = nov,
       volume = {691},
          eid = {A185},
        pages = {A185},
          doi = {10.1051/0004-6361/202348897},
archivePrefix = {arXiv},
       eprint = {2405.00232},
 primaryClass = {astro-ph.GA},
       adsurl = {https://ui.adsabs.harvard.edu/abs/2024A&A...691A.185M}
}

@ARTICLE{Navarro1997,
       author = {{Navarro}, Julio F. and {Frenk}, Carlos S. and {White}, Simon D.~M.},
        title = "{A Universal Density Profile from Hierarchical Clustering}",
      journal = {\apj},
         year = 1997,
        month = dec,
       volume = {490},
       number = {2},
        pages = {493-508},
          doi = {10.1086/304888},
archivePrefix = {arXiv},
       eprint = {astro-ph/9611107},
 primaryClass = {astro-ph},
       adsurl = {https://ui.adsabs.harvard.edu/abs/1997ApJ...490..493N}
}

@ARTICLE{Oei2022,
       author = {{Oei}, Martijn S.~S.~L. and {van Weeren}, Reinout J. and {Hardcastle}, Martin J. and {Botteon}, Andrea and {Shimwell}, Tim W. and {Dabhade}, Pratik and {Gast}, Aivin R.~D.~J.~G.~I.~B. and {R{\"o}ttgering}, Huub J.~A. and {Br{\"u}ggen}, Marcus and {Tasse}, Cyril and {Williams}, Wendy L. and {Shulevski}, Aleksandar},
        title = "{The discovery of a radio galaxy of at least 5 Mpc}",
      journal = {\aap},
         year = 2022,
        month = apr,
       volume = {660},
          eid = {A2},
        pages = {A2},
          doi = {10.1051/0004-6361/202142778},
archivePrefix = {arXiv},
       eprint = {2202.05427},
 primaryClass = {astro-ph.GA},
       adsurl = {https://ui.adsabs.harvard.edu/abs/2022A&A...660A...2O}
}

@article{2016A&A...594A..13P,
	title = "{Planck 2015 results. XIII. Cosmological parameters}",
	author = {{Planck Collaboration} and {Ade}, P.~A.~R. and {Aghanim}, N. and {Arnaud}, M. and {Ashdown}, M. and {Aumont}, J. and {Baccigalupi}, C. and {Banday}, A.~J. and {Barreiro}, R.~B. and {Bartlett}, J.~G. and {Bartolo}, N. and {Battaner}, E. and {Battye}, R. and {Benabed}, K. and {Beno{\^\i}t}, A. and {Benoit-L{\'e}vy}, A. and {Bernard}, J. -P. and {Bersanelli}, M. and {Bielewicz}, P. and {Bock}, J.~J. and {Bonaldi}, A. and {Bonavera}, L. and {Bond}, J.~R. and {Borrill}, J. and {Bouchet}, F.~R. and {Boulanger}, F. and {Bucher}, M. and {Burigana}, C. and {Butler}, R.~C. and {Calabrese}, E. and {Cardoso}, J. -F. and {Catalano}, A. and {Challinor}, A. and {Chamballu}, A. and {Chary}, R. -R. and {Chiang}, H.~C. and {Chluba}, J. and {Christensen}, P.~R. and {Church}, S. and {Clements}, D.~L. and {Colombi}, S. and {Colombo}, L.~P.~L. and {Combet}, C. and {Coulais}, A. and {Crill}, B.~P. and {Curto}, A. and {Cuttaia}, F. and {Danese}, L. and {Davies}, R.~D. and {Davis}, R.~J. and {de Bernardis}, P. and {de Rosa}, A. and {de Zotti}, G. and {Delabrouille}, J. and {D{\'e}sert}, F. -X. and {Di Valentino}, E. and {Dickinson}, C. and {Diego}, J.~M. and {Dolag}, K. and {Dole}, H. and {Donzelli}, S. and {Dor{\'e}}, O. and {Douspis}, M. and {Ducout}, A. and {Dunkley}, J. and {Dupac}, X. and {Efstathiou}, G. and {Elsner}, F. and {En{\ss}lin}, T.~A. and {Eriksen}, H.~K. and {Farhang}, M. and {Fergusson}, J. and {Finelli}, F. and {Forni}, O. and {Frailis}, M. and {Fraisse}, A.~A. and {Franceschi}, E. and {Frejsel}, A. and {Galeotta}, S. and {Galli}, S. and {Ganga}, K. and {Gauthier}, C. and {Gerbino}, M. and {Ghosh}, T. and {Giard}, M. and {Giraud-H{\'e}raud}, Y. and {Giusarma}, E. and {Gjerl{\o}w}, E. and {Gonz{\'a}lez-Nuevo}, J. and {G{\'o}rski}, K.~M. and {Gratton}, S. and {Gregorio}, A. and {Gruppuso}, A. and {Gudmundsson}, J.~E. and {Hamann}, J. and {Hansen}, F.~K. and {Hanson}, D. and {Harrison}, D.~L. and {Helou}, G. and {Henrot-Versill{\'e}}, S. and {Hern{\'a}ndez-Monteagudo}, C. and {Herranz}, D. and {Hildebrand t}, S.~R. and {Hivon}, E. and {Hobson}, M. and {Holmes}, W.~A. and {Hornstrup}, A. and {Hovest}, W. and {Huang}, Z. and {Huffenberger}, K.~M. and {Hurier}, G. and {Jaffe}, A.~H. and {Jaffe}, T.~R. and {Jones}, W.~C. and {Juvela}, M. and {Keih{\"a}nen}, E. and {Keskitalo}, R. and {Kisner}, T.~S. and {Kneissl}, R. and {Knoche}, J. and {Knox}, L. and {Kunz}, M. and {Kurki-Suonio}, H. and {Lagache}, G. and {L{\"a}hteenm{\"a}ki}, A. and {Lamarre}, J. -M. and {Lasenby}, A. and {Lattanzi}, M. and {Lawrence}, C.~R. and {Leahy}, J.~P. and {Leonardi}, R. and {Lesgourgues}, J. and {Levrier}, F. and {Lewis}, A. and {Liguori}, M. and {Lilje}, P.~B. and {Linden-V{\o}rnle}, M. and {L{\'o}pez-Caniego}, M. and {Lubin}, P.~M. and {Mac{\'\i}as-P{\'e}rez}, J.~F. and {Maggio}, G. and {Maino}, D. and {Mandolesi}, N. and {Mangilli}, A. and {Marchini}, A. and {Maris}, M. and {Martin}, P.~G. and {Martinelli}, M. and {Mart{\'\i}nez-Gonz{\'a}lez}, E. and {Masi}, S. and {Matarrese}, S. and {McGehee}, P. and {Meinhold}, P.~R. and {Melchiorri}, A. and {Melin}, J. -B. and {Mendes}, L. and {Mennella}, A. and {Migliaccio}, M. and {Millea}, M. and {Mitra}, S. and {Miville-Desch{\^e}nes}, M. -A. and {Moneti}, A. and {Montier}, L. and {Morgante}, G. and {Mortlock}, D. and {Moss}, A. and {Munshi}, D. and {Murphy}, J.~A. and {Naselsky}, P. and {Nati}, F. and {Natoli}, P. and {Netterfield}, C.~B. and {N{\o}rgaard-Nielsen}, H.~U. and {Noviello}, F. and {Novikov}, D. and {Novikov}, I. and {Oxborrow}, C.~A. and {Paci}, F. and {Pagano}, L. and {Pajot}, F. and {Paladini}, R. and {Paoletti}, D. and {Partridge}, B. and {Pasian}, F. and {Patanchon}, G. and {Pearson}, T.~J. and {Perdereau}, O. and {Perotto}, L. and {Perrotta}, F. and {Pettorino}, V. and {Piacentini}, F. and {Piat}, M. and {Pierpaoli}, E. and {Pietrobon}, D. and {Plaszczynski}, S. and {Pointecouteau}, E. and {Polenta}, G. and {Popa}, L. and {Pratt}, G.~W. and {Pr{\'e}zeau}, G. and {Prunet}, S. and {Puget}, J. -L. and {Rachen}, J.~P. and {Reach}, W.~T. and {Rebolo}, R. and {Reinecke}, M. and {Remazeilles}, M. and {Renault}, C. and {Renzi}, A. and {Ristorcelli}, I. and {Rocha}, G. and {Rosset}, C. and {Rossetti}, M. and {Roudier}, G. and {Rouill{\'e} d'Orfeuil}, B. and {Rowan-Robinson}, M. and {Rubi{\~n}o-Mart{\'\i}n}, J.~A. and {Rusholme}, B. and {Said}, N. and {Salvatelli}, V. and {Salvati}, L. and {Sandri}, M. and {Santos}, D. and {Savelainen}, M. and {Savini}, G. and {Scott}, D. and {Seiffert}, M.~D. and {Serra}, P. and {Shellard}, E.~P.~S. and {Spencer}, L.~D. and {Spinelli}, M. and {Stolyarov}, V. and {Stompor}, R. and {Sudiwala}, R. and {Sunyaev}, R. and {Sutton}, D. and {Suur-Uski}, A. -S. and {Sygnet}, J. -F. and {Tauber}, J.~A. and {Terenzi}, L. and {Toffolatti}, L. and {Tomasi}, M. and {Tristram}, M. and {Trombetti}, T. and {Tucci}, M. and {Tuovinen}, J. and {T{\"u}rler}, M. and {Umana}, G. and {Valenziano}, L. and {Valiviita}, J. and {Van Tent}, F. and {Vielva}, P. and {Villa}, F. and {Wade}, L.~A. and {Wandelt}, B.~D. and {Wehus}, I.~K. and {White}, M. and {White}, S.~D.~M. and {Wilkinson}, A. and {Yvon}, D. and {Zacchei}, A. and {Zonca}, A.},
	year = 2016,
	month = sep,
	journal = {\aap},
	volume = {594},
	pages = {A13},
	doi = {10.1051/0004-6361/201525830},
	eid = {A13},
	archivePrefix = {arXiv},
	eprint = {1502.01589},
	primaryClass = {astro-ph.CO},
	adsurl = {https://ui.adsabs.harvard.edu/abs/2016A&A...594A..13P}
}

@misc{apl,
	title = "{APLpy: Astronomical Plotting Library in Python}",
	author = {{Robitaille}, Thomas and {Bressert}, Eli},
	year = 2012,
	month = aug,
	pages = {ascl:1208.017},
	eid = {ascl:1208.017},
	archiveprefix = {ascl},
	eprint = {1208.017},
	adsurl = {https://ui.adsabs.harvard.edu/abs/2012ascl.soft08017R}
}

@ARTICLE{Sankhyayan2024,
       author = {{Sankhyayan}, Shishir and {Dabhade}, Pratik},
        title = "{Search and analysis of giant radio galaxies with associated nuclei (SAGAN). IV. Interplay with the Supercluster environment}",
      journal = {\aap},
         year = 2024,
        month = jul,
       volume = {687},
          eid = {L8},
        pages = {L8},
          doi = {10.1051/0004-6361/202450011},
archivePrefix = {arXiv},
       eprint = {2405.19154},
 primaryClass = {astro-ph.GA},
       adsurl = {https://ui.adsabs.harvard.edu/abs/2024A&A...687L...8S}
}

@article{lotssDR2,
	title = "{The LOFAR Two-metre Sky Survey. II. First data release}",
	author = {{Shimwell}, T.~W. and {Tasse}, C. and {Hardcastle}, M.~J. and {Mechev}, A.~P. and {Williams}, W.~L. and {Best}, P.~N. and {R{\"o}ttgering}, H.~J.~A. and {Callingham}, J.~R. and {Dijkema}, T.~J. and {de Gasperin}, F. and {Hoang}, D.~N. and {Hugo}, B. and {Mirmont}, M. and {Oonk}, J.~B.~R. and {Prandoni}, I. and {Rafferty}, D. and {Sabater}, J. and {Smirnov}, O. and {van Weeren}, R.~J. and {White}, G.~J. and {Atemkeng}, M. and {Bester}, L. and {Bonnassieux}, E. and {Br{\"u}ggen}, M. and {Brunetti}, G. and {Chy{\.z}y}, K.~T. and {Cochrane}, R. and {Conway}, J.~E. and {Croston}, J.~H. and {Danezi}, A. and {Duncan}, K. and {Haverkorn}, M. and {Heald}, G.~H. and {Iacobelli}, M. and {Intema}, H.~T. and {Jackson}, N. and {Jamrozy}, M. and {Jarvis}, M.~J. and {Lakhoo}, R. and {Mevius}, M. and {Miley}, G.~K. and {Morabito}, L. and {Morganti}, R. and {Nisbet}, D. and {Orr{\'u}}, E. and {Perkins}, S. and {Pizzo}, R.~F. and {Schrijvers}, C. and {Smith}, D.~J.~B. and {Vermeulen}, R. and {Wise}, M.~W. and {Alegre}, L. and {Bacon}, D.~J. and {van Bemmel}, I.~M. and {Beswick}, R.~J. and {Bonafede}, A. and {Botteon}, A. and {Bourke}, S. and {Brienza}, M. and {Calistro Rivera}, G. and {Cassano}, R. and {Clarke}, A.~O. and {Conselice}, C.~J. and {Dettmar}, R.~J. and {Drabent}, A. and {Dumba}, C. and {Emig}, K.~L. and {En{\ss}lin}, T.~A. and {Ferrari}, C. and {Garrett}, M.~A. and {G{\'e}nova-Santos}, R.~T. and {Goyal}, A. and {G{\"u}rkan}, G. and {Hale}, C. and {Harwood}, J.~J. and {Heesen}, V. and {Hoeft}, M. and {Horellou}, C. and {Jackson}, C. and {Kokotanekov}, G. and {Kondapally}, R. and {Kunert-Bajraszewska}, M. and {Mahatma}, V. and {Mahony}, E.~K. and {Mandal}, S. and {McKean}, J.~P. and {Merloni}, A. and {Mingo}, B. and {Miskolczi}, A. and {Mooney}, S. and {Nikiel-Wroczy{\'n}ski}, B. and {O'Sullivan}, S.~P. and {Quinn}, J. and {Reich}, W. and {Roskowi{\'n}ski}, C. and {Rowlinson}, A. and {Savini}, F. and {Saxena}, A. and {Schwarz}, D.~J. and {Shulevski}, A. and {Sridhar}, S.~S. and {Stacey}, H.~R. and {Urquhart}, S. and {van der Wiel}, M.~H.~D. and {Varenius}, E. and {Webster}, B. and {Wilber}, A.},
	year = 2019,
	month = feb,
	journal = {\aap},
	volume = {622},
	pages = {A1},
	doi = {10.1051/0004-6361/201833559},
	eid = {A1},
	archivePrefix = {arXiv},
	eprint = {1811.07926},
	primaryClass = {astro-ph.GA},
	adsurl = {https://ui.adsabs.harvard.edu/abs/2019A&A...622A...1S}
}

@ARTICLE{Stuardi2020,
       author = {{Stuardi}, C. and {O'Sullivan}, S.~P. and {Bonafede}, A. and {Br{\"u}ggen}, M. and {Dabhade}, P. and {Horellou}, C. and {Morganti}, R. and {Carretti}, E. and {Heald}, G. and {Iacobelli}, M. and {Vacca}, V.},
        title = "{The LOFAR view of intergalactic magnetic fields with giant radio galaxies}",
      journal = {\aap},
         year = 2020,
        month = jun,
       volume = {638},
          eid = {A48},
        pages = {A48},
          doi = {10.1051/0004-6361/202037635},
archivePrefix = {arXiv},
       eprint = {2004.05169},
 primaryClass = {astro-ph.GA},
       adsurl = {https://ui.adsabs.harvard.edu/abs/2020A&A...638A..48S}
}

@inproceedings{top05,
	title = "{TOPCAT \&amp; STIL: Starlink Table/VOTable Processing Software}",
	author = {{Taylor}, M.~B.},
	year = 2005,
	month = dec,
	booktitle = {Astronomical Data Analysis Software and Systems XIV},
	series = {Astronomical Society of the Pacific Conference Series},
	volume = {347},
	pages = {29},
	editor = {{Shopbell}, P. and {Britton}, M. and {Ebert}, R.},
	adsurl = {https://ui.adsabs.harvard.edu/abs/2005ASPC..347...29T}
}

@ARTICLE{Tanimura2020,
       author = {{Tanimura}, H. and {Aghanim}, N. and {Bonjean}, V. and {Malavasi}, N. and {Douspis}, M.},
        title = "{Density and temperature of cosmic-web filaments on scales of tens of megaparsecs}",
      journal = {\aap},
         year = 2020,
        month = may,
       volume = {637},
          eid = {A41},
        pages = {A41},
          doi = {10.1051/0004-6361/201937158},
archivePrefix = {arXiv},
       eprint = {1911.09706},
 primaryClass = {astro-ph.CO},
       adsurl = {https://ui.adsabs.harvard.edu/abs/2020A&A...637A..41T}
}

@ARTICLE{Tempel2014_fil_cat,
       author = {{Tempel}, E. and {Stoica}, R.~S. and {Mart{\'\i}nez}, V.~J. and {Liivam{\"a}gi}, L.~J. and {Castellan}, G. and {Saar}, E.},
        title = "{Detecting filamentary pattern in the cosmic web: a catalogue of filaments for the SDSS}",
      journal = {\mnras},
         year = 2014,
        month = mar,
       volume = {438},
       number = {4},
        pages = {3465-3482},
          doi = {10.1093/mnras/stt2454},
archivePrefix = {arXiv},
       eprint = {1308.2533},
 primaryClass = {astro-ph.CO},
       adsurl = {https://ui.adsabs.harvard.edu/abs/2014MNRAS.438.3465T}
}

@ARTICLE{Tempel_Bisous2016,
       author = {{Tempel}, E. and {Stoica}, R.~S. and {Kipper}, R. and {Saar}, E.},
        title = "{Bisous model-Detecting filamentary patterns in point processes}",
      journal = {Astronomy and Computing},
         year = 2016,
        month = jul,
       volume = {16},
        pages = {17-25},
          doi = {10.1016/j.ascom.2016.03.004},
archivePrefix = {arXiv},
       eprint = {1603.08957},
 primaryClass = {astro-ph.CO},
       adsurl = {https://ui.adsabs.harvard.edu/abs/2016A&C....16...17T}
}

@ARTICLE{Wang2024,
       author = {{Wang}, Wei and {Wang}, Peng and {Guo}, Hong and {Kang}, Xi and {Libeskind}, Noam I. and {Gal{\'a}rraga-Espinosa}, Daniela and {Springel}, Volker and {Kannan}, Rahul and {Hernquist}, Lars and {Pakmor}, R{\"u}diger and {Yu}, Hao-Ran and {Bose}, Sownak and {Guo}, Quan and {Yu}, Luo and {Hern{\'a}ndez-Aguayo}, C{\'e}sar},
        title = "{The boundary of cosmic filaments}",
      journal = {\mnras},
         year = 2024,
        month = aug,
       volume = {532},
       number = {4},
        pages = {4604-4615},
          doi = {10.1093/mnras/stae1801},
archivePrefix = {arXiv},
       eprint = {2402.11678},
 primaryClass = {astro-ph.CO},
       adsurl = {https://ui.adsabs.harvard.edu/abs/2024MNRAS.532.4604W}
}

@ARTICLE{Willis1974,
       author = {{Willis}, A.~G. and {Strom}, R.~G. and {Wilson}, A.~S.},
        title = "{3C236, DA240; the largest radio sources known}",
      journal = {\nat},
         year = 1974,
        month = aug,
       volume = {250},
       number = {5468},
        pages = {625-630},
          doi = {10.1038/250625a0},
       adsurl = {https://ui.adsabs.harvard.edu/abs/1974Natur.250..625W}
}

@INBOOK{wita-grg-agn,
	title = "{Giant Radio Galaxies via Inverse Compton Weakened Jets}",
	author = {{Wiita}, Paul J. and {Rosen}, Alexander and {Gopal-Krishna} and {Saripalli}, L.},
	year = 1989,
	booktitle = {Hot Spots in Extragalactic Radio Sources},
	volume = {327},
	pages = {173},
	doi = {10.1007/BFb0036027},
	editor = {{Meisenheimer}, Klaus and {Roeser}, Hermann-Josef},
	adsurl = {https://ui.adsabs.harvard.edu/abs/1989LNP...327..173W}
}

@ARTICLE{Wang2021,
       author = {{Wang}, Peng and {Libeskind}, Noam I. and {Tempel}, Elmo and {Kang}, Xi and {Guo}, Quan},
        title = "{Possible observational evidence for cosmic filament spin}",
      journal = {Nature Astronomy},
         year = 2021,
        month = aug,
       volume = {5},
        pages = {839-845},
          doi = {10.1038/s41550-021-01380-6},
archivePrefix = {arXiv},
       eprint = {2106.05989},
 primaryClass = {astro-ph.GA},
       adsurl = {https://ui.adsabs.harvard.edu/abs/2021NatAs...5..839W}
}

@ARTICLE{Xia2021,
       author = {{Xia}, Qianli and {Neyrinck}, Mark C. and {Cai}, Yan-Chuan and {Arag{\'o}n-Calvo}, Miguel A.},
        title = "{Intergalactic filaments spin}",
      journal = {\mnras},
         year = 2021,
        month = sep,
       volume = {506},
       number = {1},
        pages = {1059-1072},
          doi = {10.1093/mnras/stab1713},
archivePrefix = {arXiv},
       eprint = {2006.02418},
 primaryClass = {astro-ph.CO},
       adsurl = {https://ui.adsabs.harvard.edu/abs/2021MNRAS.506.1059X}
}

@article{sdssyork,
	title = {{The Sloan Digital Sky Survey: Technical Summary}},
	author = {{York}, D.~G. and {Adelman}, J. and {Anderson}, Jr., J.~E. and {Anderson}, S.~F. and {Annis}, J. and {Bahcall}, N.~A. and {Bakken}, J.~A. and {Barkhouser}, R. and {Bastian}, S. and {Berman}, E. and {Boroski}, W.~N. and {Bracker}, S. and {Briegel}, C. and {Briggs}, J.~W. and {Brinkmann}, J. and {Brunner}, R. and {Burles}, S. and {Carey}, L. and {Carr}, M.~A. and {Castander}, F.~J. and {Chen}, B. and {Colestock}, P.~L. and {Connolly}, A.~J. and {Crocker}, J.~H. and {Csabai}, I. and {Czarapata}, P.~C. and {Davis}, J.~E. and {Doi}, M. and {Dombeck}, T. and {Eisenstein}, D. and {Ellman}, N. and {Elms}, B.~R. and {Evans}, M.~L. and {Fan}, X. and {Federwitz}, G.~R. and {Fiscelli}, L. and {Friedman}, S. and {Frieman}, J.~A. and {Fukugita}, M. and {Gillespie}, B. and {Gunn}, J.~E. and {Gurbani}, V.~K. and {de Haas}, E. and {Haldeman}, M. and {Harris}, F.~H. and {Hayes}, J. and {Heckman}, T.~M. and {Hennessy}, G.~S. and {Hindsley}, R.~B. and {Holm}, S. and {Holmgren}, D.~J. and {Huang}, C.-h. and {Hull}, C. and {Husby}, D. and {Ichikawa}, S.-I. and {Ichikawa}, T. and {Ivezi{\'c}}, {\v Z}. and {Kent}, S. and {Kim}, R.~S.~J. and {Kinney}, E. and {Klaene}, M. and {Kleinman}, A.~N. and {Kleinman}, S. and {Knapp}, G.~R. and {Korienek}, J. and {Kron}, R.~G. and {Kunszt}, P.~Z. and {Lamb}, D.~Q. and {Lee}, B. and {Leger}, R.~F. and {Limmongkol}, S. and {Lindenmeyer}, C. and {Long}, D.~C. and {Loomis}, C. and {Loveday}, J. and {Lucinio}, R. and {Lupton}, R.~H. and {MacKinnon}, B. and {Mannery}, E.~J. and {Mantsch}, P.~M. and {Margon}, B. and {McGehee}, P. and {McKay}, T.~A. and {Meiksin}, A. and {Merelli}, A. and {Monet}, D.~G. and {Munn}, J.~A. and {Narayanan}, V.~K. and {Nash}, T. and {Neilsen}, E. and {Neswold}, R. and {Newberg}, H.~J. and {Nichol}, R.~C. and {Nicinski}, T. and {Nonino}, M. and {Okada}, N. and {Okamura}, S. and {Ostriker}, J.~P. and {Owen}, R. and {Pauls}, A.~G. and {Peoples}, J. and {Peterson}, R.~L. and {Petravick}, D. and {Pier}, J.~R. and {Pope}, A. and {Pordes}, R. and {Prosapio}, A. and {Rechenmacher}, R. and {Quinn}, T.~R. and {Richards}, G.~T. and {Richmond}, M.~W. and {Rivetta}, C.~H. and {Rockosi}, C.~M. and {Ruthmansdorfer}, K. and {Sandford}, D. and {Schlegel}, D.~J. and {Schneider}, D.~P. and {Sekiguchi}, M. and {Sergey}, G. and {Shimasaku}, K. and {Siegmund}, W.~A. and {Smee}, S. and {Smith}, J.~A. and {Snedden}, S. and {Stone}, R. and {Stoughton}, C. and {Strauss}, M.~A. and {Stubbs}, C. and {SubbaRao}, M. and {Szalay}, A.~S. and {Szapudi}, I. and {Szokoly}, G.~P. and {Thakar}, A.~R. and {Tremonti}, C. and {Tucker}, D.~L. and {Uomoto}, A. and {Vanden Berk}, D. and {Vogeley}, M.~S. and {Waddell}, P. and {Wang}, S.-i. and {Watanabe}, M. and {Weinberg}, D.~H. and {Yanny}, B. and {Yasuda}, N. and {SDSS Collaboration}},
	year = 2000,
	month = sep,
	journal = {\aj},
	volume = 120,
	pages = {1579--1587},
	doi = {10.1086/301513},
	adsurl = {http://adsabs.harvard.edu/abs/2000AJ....120.1579Y},
	eprint = {astro-ph/0006396}
}

@ARTICLE{MassiveBlack-II,
       author = {{Khandai}, Nishikanta and {Di Matteo}, Tiziana and {Croft}, Rupert and {Wilkins}, Stephen and {Feng}, Yu and {Tucker}, Evan and {DeGraf}, Colin and {Liu}, Mao-Sheng},
        title = "{The MassiveBlack-II simulation: the evolution of haloes and galaxies to z {\ensuremath{\sim}} 0}",
      journal = {\mnras},
         year = 2015,
        month = jun,
       volume = {450},
       number = {2},
        pages = {1349-1374},
          doi = {10.1093/mnras/stv627},
archivePrefix = {arXiv},
       eprint = {1402.0888},
 primaryClass = {astro-ph.CO},
       adsurl = {https://ui.adsabs.harvard.edu/abs/2015MNRAS.450.1349K}
}

@ARTICLE{WH15,
       author = {{Wen}, Z.~L. and {Han}, J.~L.},
        title = "{Calibration of the Optical Mass Proxy for Clusters of Galaxies and an Update of the WHL12 Cluster Catalog}",
      journal = {\apj},
         year = 2015,
        month = jul,
       volume = {807},
       number = {2},
          eid = {178},
        pages = {178},
          doi = {10.1088/0004-637X/807/2/178},
archivePrefix = {arXiv},
       eprint = {1506.04503},
 primaryClass = {astro-ph.GA},
       adsurl = {https://ui.adsabs.harvard.edu/abs/2015ApJ...807..178W}
}

@ARTICLE{de_Jong19,
       author = {{de Jong}, R.~S. and {Agertz}, O. and {Berbel}, A.~A. and {Aird}, J. and {Alexander}, D.~A. and {Amarsi}, A. and {Anders}, F. and {Andrae}, R. and {Ansarinejad}, B. and {Ansorge}, W. and {Antilogus}, P. and {Anwand-Heerwart}, H. and {Arentsen}, A. and {Arnadottir}, A. and {Asplund}, M. and {Auger}, M. and {Azais}, N. and {Baade}, D. and {Baker}, G. and {Baker}, S. and {Balbinot}, E. and {Baldry}, I.~K. and {Banerji}, M. and {Barden}, S. and {Barklem}, P. and {Barth{\'e}l{\'e}my-Mazot}, E. and {Battistini}, C. and {Bauer}, S. and {Bell}, C.~P.~M. and {Bellido-Tirado}, O. and {Bellstedt}, S. and {Belokurov}, V. and {Bensby}, T. and {Bergemann}, M. and {Bestenlehner}, J.~M. and {Bielby}, R. and {Bilicki}, M. and {Blake}, C. and {Bland-Hawthorn}, J. and {Boeche}, C. and {Boland}, W. and {Boller}, T. and {Bongard}, S. and {Bongiorno}, A. and {Bonifacio}, P. and {Boudon}, D. and {Brooks}, D. and {Brown}, M.~J.~I. and {Brown}, R. and {Br{\"u}ggen}, M. and {Brynnel}, J. and {Brzeski}, J. and {Buchert}, T. and {Buschkamp}, P. and {Caffau}, E. and {Caillier}, P. and {Carrick}, J. and {Casagrande}, L. and {Case}, S. and {Casey}, A. and {Cesarini}, I. and {Cescutti}, G. and {Chapuis}, D. and {Chiappini}, C. and {Childress}, M. and {Christlieb}, N. and {Church}, R. and {Cioni}, M. -R.~L. and {Cluver}, M. and {Colless}, M. and {Collett}, T. and {Comparat}, J. and {Cooper}, A. and {Couch}, W. and {Courbin}, F. and {Croom}, S. and {Croton}, D. and {Daguis{\'e}}, E. and {Dalton}, G. and {Davies}, L.~J.~M. and {Davis}, T. and {de Laverny}, P. and {Deason}, A. and {Dionies}, F. and {Disseau}, K. and {Doel}, P. and {D{\"o}scher}, D. and {Driver}, S.~P. and {Dwelly}, T. and {Eckert}, D. and {Edge}, A. and {Edvardsson}, B. and {Youssoufi}, D.~E. and {Elhaddad}, A. and {Enke}, H. and {Erfanianfar}, G. and {Farrell}, T. and {Fechner}, T. and {Feiz}, C. and {Feltzing}, S. and {Ferreras}, I. and {Feuerstein}, D. and {Feuillet}, D. and {Finoguenov}, A. and {Ford}, D. and {Fotopoulou}, S. and {Fouesneau}, M. and {Frenk}, C. and {Frey}, S. and {Gaessler}, W. and {Geier}, S. and {Gentile Fusillo}, N. and {Gerhard}, O. and {Giannantonio}, T. and {Giannone}, D. and {Gibson}, B. and {Gillingham}, P. and {Gonz{\'a}lez-Fern{\'a}ndez}, C. and {Gonzalez-Solares}, E. and {Gottloeber}, S. and {Gould}, A. and {Grebel}, E.~K. and {Gueguen}, A. and {Guiglion}, G. and {Haehnelt}, M. and {Hahn}, T. and {Hansen}, C.~J. and {Hartman}, H. and {Hauptner}, K. and {Hawkins}, K. and {Haynes}, D. and {Haynes}, R. and {Heiter}, U. and {Helmi}, A. and {Aguayo}, C.~H. and {Hewett}, P. and {Hinton}, S. and {Hobbs}, D. and {Hoenig}, S. and {Hofman}, D. and {Hook}, I. and {Hopgood}, J. and {Hopkins}, A. and {Hourihane}, A. and {Howes}, L. and {Howlett}, C. and {Huet}, T. and {Irwin}, M. and {Iwert}, O. and {Jablonka}, P. and {Jahn}, T. and {Jahnke}, K. and {Jarno}, A. and {Jin}, S. and {Jofre}, P. and {Johl}, D. and {Jones}, D. and {J{\"o}nsson}, H. and {Jordan}, C. and {Karovicova}, I. and {Khalatyan}, A. and {Kelz}, A. and {Kennicutt}, R. and {King}, D. and {Kitaura}, F. and {Klar}, J. and {Klauser}, U. and {Kneib}, J. -P. and {Koch}, A. and {Koposov}, S. and {Kordopatis}, G. and {Korn}, A. and {Kosmalski}, J. and {Kotak}, R. and {Kovalev}, M. and {Kreckel}, K. and {Kripak}, Y. and {Krumpe}, M. and {Kuijken}, K. and {Kunder}, A. and {Kushniruk}, I. and {Lam}, M.~I. and {Lamer}, G. and {Laurent}, F. and {Lawrence}, J. and {Lehmitz}, M. and {Lemasle}, B. and {Lewis}, J. and {Li}, B. and {Lidman}, C. and {Lind}, K. and {Liske}, J. and {Lizon}, J. -L. and {Loveday}, J. and {Ludwig}, H. -G. and {McDermid}, R.~M. and {Maguire}, K. and {Mainieri}, V. and {Mali}, S. and {Mandel}, H. and {Mandel}, K. and {Mannering}, L. and {Martell}, S. and {Martinez Delgado}, D. and {Matijevic}, G. and {McGregor}, H. and {McMahon}, R. and {McMillan}, P. and {Mena}, O. and {Merloni}, A. and {Meyer}, M.~J. and {Michel}, C. and {Micheva}, G. and {Migniau}, J. -E. and {Minchev}, I. and {Monari}, G. and {Muller}, R. and {Murphy}, D. and {Muthukrishna}, D. and {Nandra}, K. and {Navarro}, R. and {Ness}, M. and {Nichani}, V. and {Nichol}, R. and {Nicklas}, H. and {Niederhofer}, F. and {Norberg}, P. and {Obreschkow}, D. and {Oliver}, S. and {Owers}, M. and {Pai}, N. and {Pankratow}, S. and {Parkinson}, D. and {Paschke}, J. and {Paterson}, R. and {Pecontal}, A. and {Parry}, I. and {Phillips}, D. and {Pillepich}, A. and {Pinard}, L. and {Pirard}, J. and {Piskunov}, N. and {Plank}, V. and {Pl{\"u}schke}, D. and {Pons}, E. and {Popesso}, P. and {Power}, C. and {Pragt}, J. and {Pramskiy}, A. and {Pryer}, D. and {Quattri}, M. and {Queiroz}, A.~B. d. A. and {Quirrenbach}, A. and {Rahurkar}, S. and {Raichoor}, A. and {Ramstedt}, S. and {Rau}, A. and {Recio-Blanco}, A. and {Reiss}, R. and {Renaud}, F. and {Revaz}, Y. and {Rhode}, P. and {Richard}, J. and {Richter}, A.~D. and {Rix}, H. -W. and {Robotham}, A.~S.~G. and {Roelfsema}, R. and {Romaniello}, M. and {Rosario}, D. and {Rothmaier}, F. and {Roukema}, B. and {Ruchti}, G. and {Rupprecht}, G. and {Rybizki}, J. and {Ryde}, N. and {Saar}, A. and {Sadler}, E. and {Sahl{\'e}n}, M. and {Salvato}, M. and {Sassolas}, B. and {Saunders}, W. and {Saviauk}, A. and {Sbordone}, L. and {Schmidt}, T. and {Schnurr}, O. and {Scholz}, R. -D. and {Schwope}, A. and {Seifert}, W. and {Shanks}, T. and {Sheinis}, A. and {Sivov}, T. and {Sk{\'u}lad{\'o}ttir}, {\'A}. and {Smartt}, S. and {Smedley}, S. and {Smith}, G. and {Smith}, R. and {Sorce}, J. and {Spitler}, L. and {Starkenburg}, E. and {Steinmetz}, M. and {Stilz}, I. and {Storm}, J. and {Sullivan}, M. and {Sutherland}, W. and {Swann}, E. and {Tamone}, A. and {Taylor}, E.~N. and {Teillon}, J. and {Tempel}, E. and {ter Horst}, R. and {Thi}, W. -F. and {Tolstoy}, E. and {Trager}, S. and {Traven}, G. and {Tremblay}, P. -E. and {Tresse}, L. and {Valentini}, M. and {van de Weygaert}, R. and {van den Ancker}, M. and {Veljanoski}, J. and {Venkatesan}, S. and {Wagner}, L. and {Wagner}, K. and {Walcher}, C.~J. and {Waller}, L. and {Walton}, N. and {Wang}, L. and {Winkler}, R. and {Wisotzki}, L. and {Worley}, C.~C. and {Worseck}, G. and {Xiang}, M. and {Xu}, W. and {Yong}, D. and {Zhao}, C. and {Zheng}, J. and {Zscheyge}, F. and {Zucker}, D.},
        title = "{4MOST: Project overview and information for the First Call for Proposals}",
      journal = {The Messenger},
         year = 2019,
        month = mar,
       volume = {175},
        pages = {3-11},
          doi = {10.18727/0722-6691/5117},
archivePrefix = {arXiv},
       eprint = {1903.02464},
 primaryClass = {astro-ph.IM},
       adsurl = {https://ui.adsabs.harvard.edu/abs/2019Msngr.175....3D}
}

@ARTICLE{AGN_unification,
       author = {{Urry}, C. Megan and {Padovani}, Paolo},
        title = "{Unified Schemes for Radio-Loud Active Galactic Nuclei}",
      journal = {\pasp},
         year = 1995,
        month = sep,
       volume = {107},
        pages = {803},
          doi = {10.1086/133630},
archivePrefix = {arXiv},
       eprint = {astro-ph/9506063},
 primaryClass = {astro-ph},
       adsurl = {https://ui.adsabs.harvard.edu/abs/1995PASP..107..803U}
}

@ARTICLE{Condon1997,
       author = {{Condon}, J.~J.},
        title = "{Errors in Elliptical Gaussian Fits}",
      journal = {\pasp},
         year = 1997,
        month = feb,
       volume = {109},
        pages = {166-172},
          doi = {10.1086/133871},
       adsurl = {https://ui.adsabs.harvard.edu/abs/1997PASP..109..166C}
}

@INPROCEEDINGS{Fomalont1999,
       author = {{Fomalont}, Ed B.},
        title = "{Image Analysis}",
    booktitle = {Synthesis Imaging in Radio Astronomy II},
         year = 1999,
       editor = {{Taylor}, G.~B. and {Carilli}, C.~L. and {Perley}, R.~A.},
       series = {Astronomical Society of the Pacific Conference Series},
       volume = {180},
        month = jan,
        pages = {301},
       adsurl = {https://ui.adsabs.harvard.edu/abs/1999ASPC..180..301F}
}

\begin{appendix}

\onecolumn
\section{Images}\label{sec:appendix-im}

Here, we present optical RGB images from the Legacy survey overlaid with radio contours from radio surveys, along with filaments outlined.
\begin{figure*}
\includegraphics[scale=0.68]{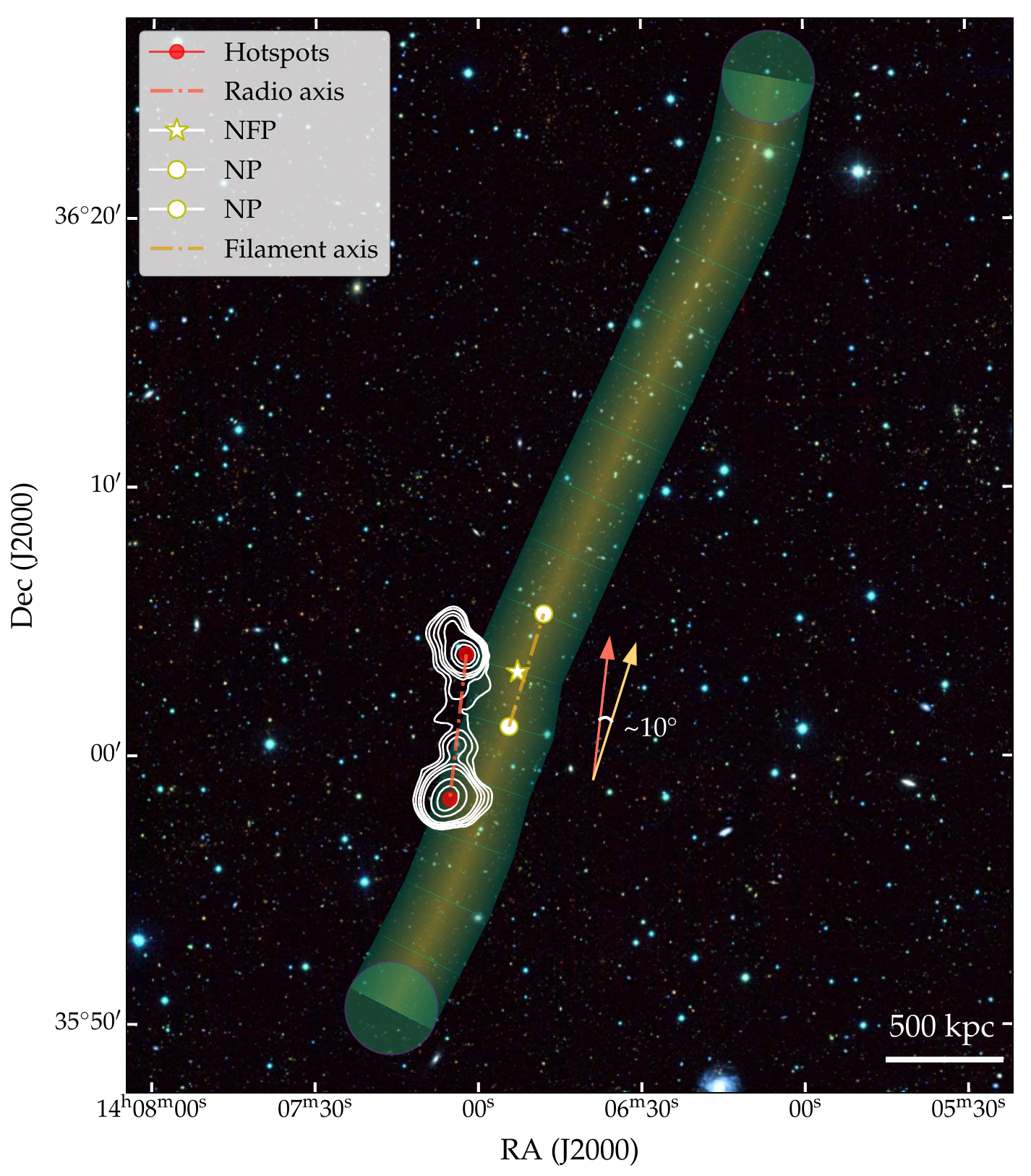}
\caption{The figure gives the view of GRG (RA: 211.7667$^{\circ}$, Dec: 36.0082$^{\circ}$), whose jet axis is nearly parallel to the filament, forming an angle of $\sim$10$^{\circ}$ with the filament axis. A green cylindrical spine traces the inner
400~kpc of the nearest filament; the nearest filament point (NFP) is indicated by a yellow star, and neighbouring points (NP) are shown as white circles. The white NVSS (beam $=$\,45$\arcsec$) 1400 MHz radio contours above 3$\sigma$ ($\sigma \sim 0.45$ mJy beam$^{-1}$) are overlaid on the Legacy Surveys optical RGB image. The hotspots are marked by red circles. The red and yellow dashed lines indicate the radio and filament axes, respectively. Arrows of the same colours are drawn parallel to their axes. The angle between the two arrows quantifies the alignment of the GRG with the filament on the plane of the sky.}
\label{fig:Fil_GRG_10}
\end{figure*} 
\begin{figure*}
\includegraphics[scale=0.68]{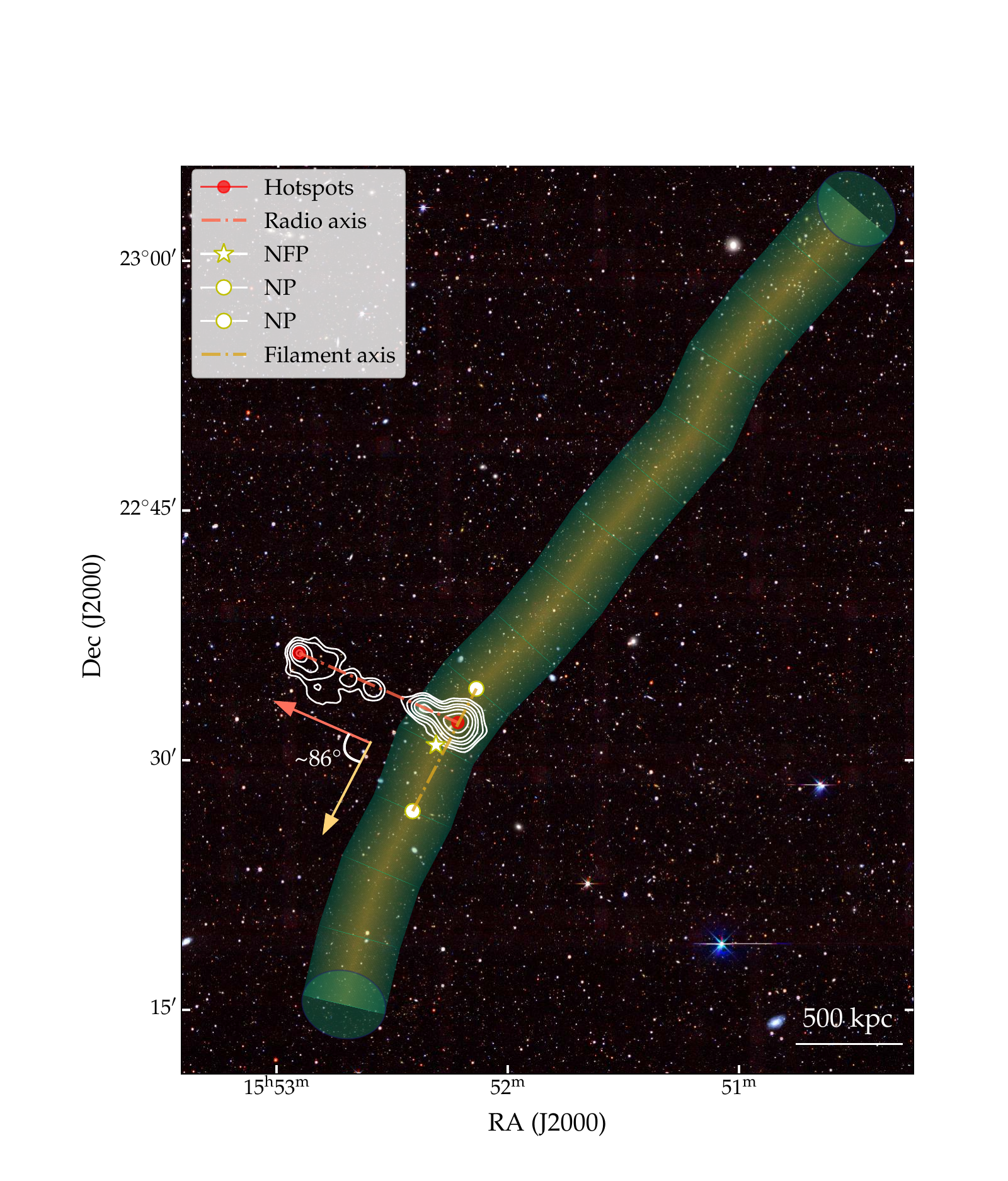}
\caption{Same plotting conventions as in Fig.~\ref{fig:Fil_GRG_10}. This figure shows the GRG at RA: 238.0931$^{\circ}$, Dec: 22.5533$^{\circ}$. White NVSS 1400 MHz contours (beam = 45$\arcsec$, $\sigma \sim 0.45$ mJy beam$^{-1}$) above 3$\sigma$ are overlaid on the Legacy Surveys RGB image. The nearest filament is traced over its inner 400 kpc, the yellow star marks the NFP with neighbouring points (NP) indicated by white circles. The GRG jet axis is nearly perpendicular to the filament, forming an angle of $\sim$86$^{\circ}$ with the filament axis. Red circles denote the hotspot positions. The radio axis and filament axis are shown by red and yellow dashed lines, respectively, with the same coloured arrows drawn parallel to each. The angle between the two arrows measures the alignment of the GRG with the filament on the plane of the sky.}
\label{fig:Fil_GRG_86}
\end{figure*} 
\begin{figure*}
\includegraphics[width=\textwidth]{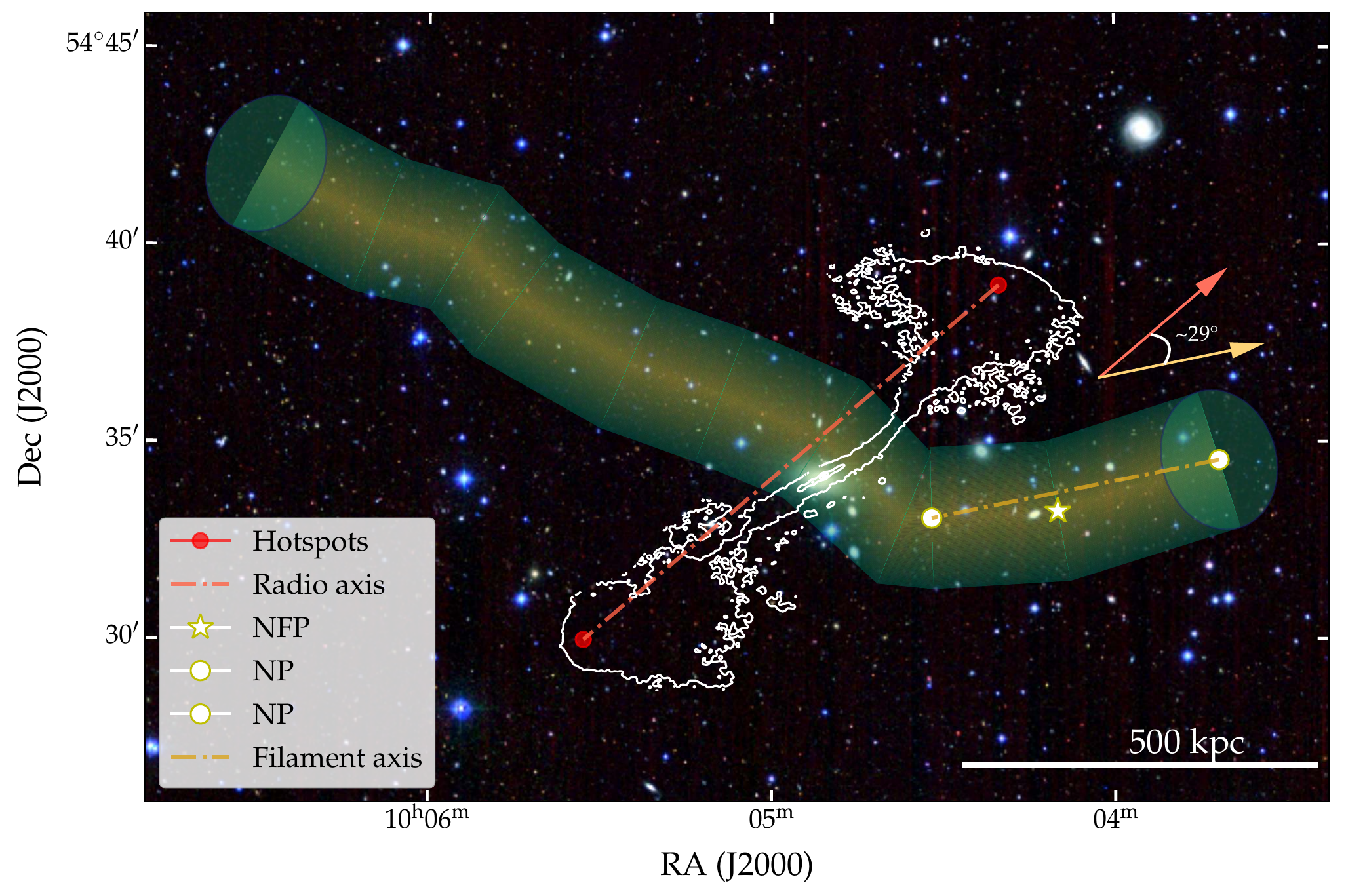}
\caption{As in Fig.~\ref{fig:Fil_GRG_10} but for the GRG at RA: 151.2168$^{\circ}$, Dec: 54.5678$^{\circ}$. Here, the radio map is from LoTSS at 144 MHz with a 6$\arcsec$ beam, shown as white contours above 3$\sigma$ ($\sigma \sim50$ $\mu$Jy beam$^{-1}$). The filament spine is traced over its inner 200 kpc, the yellow star marks the NFP and its neighbouring points (NP) are plotted as white circles. The GRG jet axis forms an angle of $\sim$29$^{\circ}$ with the filament axis. Red circles mark the hotspot positions. The background is the Legacy Surveys optical RGB image of the sky. The dashed lines indicate the radio (red) and filament (yellow) axes; corresponding red and yellow arrows are drawn parallel to them. The angle between the arrows represents the GRG-filament alignment on the plane of the sky.}
\label{fig:Fil_GRG1}
\end{figure*} 

\newpage
\onecolumn
\section{Table} \label{sec:appendix-tab}
As described in Sec.~\ref{distance}, here we list the samples of GRGs and SRGs with their properties.

\setlength{\tabcolsep}{5.0pt}
\begin{longtable}{lllclllllllr}
\captionsetup{width=\textwidth}
\caption{Properties of FRII GRGs and SRGs in cosmic web filaments. Columns (2) \& (3) represent the right ascension (RA) and declination (Dec) in degrees of the host galaxies of the GRGs and SRGs. Column (4) lists the spectroscopic redshifts of the hosts. Column (5) lists the projected linear sizes of the sources in Mpc. Columns (6), (7), (8) \& (9) respectively, state the RA and Dec of two hotspots. Columns (10) \& (11) contain the 3D distances of GRGs and SRGs from their local filaments, and their orientations with them. Column (12) gives the measurements of their arm-length ratios.} \label{tab:main} \\

\hline
 \rule{0pt}{10pt}No &       RA &      Dec &       $z$ &  Size &  RA: HS1 & Dec: HS1 &  RA: HS2 & Dec: HS2 &  D$\rm _{fil}$ & Alignment &  ALR \\
 \rule{0pt}{10pt}& (deg) & (deg) &      & (Mpc) & (deg) & (deg) & (deg) & (deg) & (Mpc) &  (deg) &   \\
\rule{0pt}{10pt}(1) & (2) & (3) & (4) & (5) & (6) & (7) & (8) & (9) & (10) & (11) & (12)  \\
\hline
\endfirsthead
\caption{continued.}\\
\hline
\hline
\rule{0pt}{10pt} No &       RA &      Dec &       $z$ &  Size &  RA: HS1 & Dec: HS1 &  RA: HS2 & Dec: HS2 &  D$\rm _{fil}$ & Alignment &  ALR \\
\rule{0pt}{10pt} & (deg) & (deg) &      & (Mpc) & (deg) & (deg) & (deg) & (deg) & (Mpc) &  (deg) &   \\
\rule{0pt}{10pt}(1) & (2) & (3) & (4) & (5) & (6) & (7) & (8) & (9) & (10) & (11) & (12) \\
\hline
\endhead
\hline
\endfoot
 \rule{0pt}{10pt}&       &       &      &    &       &   GRG    &       &       &    &        &   \\
\hline
  1 & 116.0338 &  43.9917 & 0.13484 &  0.73 & 116.0360 &  44.0212 & 116.0240 &  43.9600 &   1.5 &      84.2 & 0.91 \\
  2 & 118.1446 &  35.8398 & 0.13650 &  2.34 & 118.1587 &  35.8580 & 118.1369 &  35.8150 &   3.2 &      54.0 & 0.84 \\
  3 & 119.1594 &  32.4631 & 0.14620 &  1.29 & 119.2373 &  32.4730 & 119.1061 &  32.4435 &   0.8 &      89.8 & 0.74 \\
  4 & 127.8645 &  32.3241 & 0.05120 &  0.79 & 127.8243 &  32.3065 & 127.9014 &  32.3418 &   2.2 &      47.3 & 0.94 \\
  5 & 127.9987 &  30.6585 & 0.10704 &  2.28 & 128.0388 &  30.7812 & 127.9561 &  30.4930 &   0.8 &      74.3 & 0.75 \\
  6 & 129.0320 &  26.8121 & 0.08780 &  1.00 & 129.1003 &  26.8027 & 128.9730 &  26.8143 &   0.5 &      23.7 & 0.85 \\
  7 & 131.2433 &  42.0774 & 0.14932 &  1.70 & 131.2454 &  42.0785 & 131.2399 &  42.0759 &   0.2 &      80.9 & 0.64 \\
  8 & 131.3629 &  44.9240 & 0.15062 &  0.81 & 131.3403 &  44.9638 & 131.3691 &  44.9022 &   0.4 &      46.7 & 0.52 \\
  9 & 134.8236 &  63.7514 & 0.12010 &  0.83 & 134.8774 &  63.7794 & 134.7889 &  63.7018 &   0.8 &      61.4 & 0.71 \\
 10 & 139.9519 &  57.8489 & 0.13695 &  0.72 & 139.9678 &  57.8496 & 139.9303 &  57.8535 &   1.1 &      12.4 & 0.69 \\
 11 & 140.8813 &  24.4463 & 0.03452 &  0.77 & 140.8565 &  24.5825 & 140.9423 &  24.3533 &   2.5 &      58.4 & 0.79 \\
 12 & 143.5899 &  38.3849 & 0.12299 &  0.72 & 143.5735 &  38.4254 & 143.6050 &  38.3559 &   0.4 &      67.5 & 0.74 \\
 13 & 144.6027 &  64.7014 & 0.13881 &  1.06 & 144.5699 &  64.6916 & 144.6394 &  64.7099 &   0.2 &      49.2 & 0.96 \\
 14 & 146.7867 &  42.1904 & 0.07213 &  0.81 & 146.7623 &  42.1707 & 146.8467 &  42.2347 &   1.7 &      38.5 & 0.43 \\
 15 & 149.9183 &  17.4245 & 0.12347 &  1.21 & 149.9442 &  17.4969 & 149.8784 &  17.3565 &   0.3 &      44.3 & 0.98 \\
 16 & 151.2168 &  54.5678 & 0.04701 &  0.89 & 151.0845 &  54.6498 & 151.3869 &  54.5000 &   0.4 &      29.1 & 0.94 \\
 17 & 151.5073 &  34.9029 & 0.09936 &  4.88 & 151.9172 &  34.6816 & 151.2578 &  35.0336 &   1.3 &      78.1 & 0.60 \\
 18 & 151.5984 &  24.0906 & 0.07517 &  0.92 & 151.5922 &  24.1003 & 151.5968 &  24.0739 &   1.2 &       6.5 & 0.67 \\
 19 & 154.8433 &  32.1561 & 0.09583 &  0.72 & 154.9022 &  32.1474 & 154.7916 &  32.1669 &   1.3 &      75.9 & 0.89 \\
 20 & 155.3509 &  12.2848 & 0.12938 &  2.32 & 155.4510 &  12.3291 & 155.2553 &  12.2072 &   4.5 &      49.4 & 0.88 \\
 21 & 156.3216 &  48.5371 & 0.14920 &  1.36 & 156.2364 &  48.5520 & 156.4044 &  48.5245 &   2.3 &      60.3 & 0.97 \\
 22 & 158.0586 &  27.9333 & 0.08540 &  1.09 & 158.0514 &  27.9926 & 158.0719 &  27.8265 &   1.0 &      14.8 & 0.55 \\
 23 & 162.7187 &  40.0141 & 0.12947 &  0.95 & 162.6950 &  40.0150 & 162.7521 &  40.0152 &   2.6 &      40.6 & 0.71 \\
 24 & 169.9331 &  13.9639 & 0.06874 &  1.14 & 169.8981 &  14.0032 & 169.9755 &  13.9452 &   0.8 &      64.4 & 0.87 \\
 25 & 171.8767 &  17.1266 & 0.12973 &  1.15 & 171.8518 &  17.1310 & 171.9284 &  17.1218 &   0.3 &      84.7 & 0.49 \\
 26 & 176.1133 &  37.1422 & 0.11482 &  0.86 & 176.0804 &  37.1172 & 176.1442 &  37.1714 &   1.2 &      82.8 & 0.95 \\
 27 & 176.8420 &  35.0187 & 0.06289 &  0.91 & 176.9539 &  34.9909 & 176.7518 &  35.0280 &   1.4 &      68.0 & 0.78 \\
 28 & 183.8924 &  13.9430 & 0.09341 &  0.82 & 183.8401 &  13.9322 & 183.9442 &  13.9429 &   1.1 &      50.8 & 0.97 \\
 29 & 186.5938 &  64.1061 & 0.11024 &  0.88 & 186.5809 &  64.0928 & 186.6699 &  64.1397 &   2.3 &      70.9 & 0.31 \\
 30 & 189.4413 &  -1.2378 & 0.13531 &  1.72 & 189.4322 &  -1.1382 & 189.4434 &  -1.2673 &   0.3 &      22.1 & 0.30 \\
 31 & 198.0707 &  44.8393 & 0.03580 &  1.00 & 198.3296 &  44.8310 & 197.7798 &  44.8510 &   1.8 &      78.3 & 0.89 \\
 32 & 202.1432 &  -3.1291 & 0.08525 &  1.33 & 202.1944 &  -3.0214 & 202.1051 &  -3.2260 &   2.2 &      56.7 & 0.87 \\
 33 & 211.7667 &  36.0082 & 0.10475 &  0.93 & 211.7593 &  36.0632 & 211.7718 &  35.9735 &   0.5 &      10.7 & 0.63 \\
 34 & 217.0800 &  29.3123 & 0.08699 &  1.72 & 217.1680 &  29.3891 & 216.9565 &  29.2190 &   1.6 &      79.1 & 0.76 \\
 35 & 218.2409 &   4.6209 & 0.10552 &  0.79 & 218.2259 &   4.5972 & 218.2817 &   4.6588 &   2.0 &      38.4 & 0.50 \\
 36 & 221.3642 &   9.5383 & 0.09448 &  0.78 & 221.3243 &   9.5339 & 221.4049 &   9.5409 &   1.7 &      38.5 & 0.98 \\
 37 & 223.6808 &  12.4196 & 0.12199 &  0.94 & 223.6597 &  12.4347 & 223.7185 &  12.4034 &   1.3 &      17.6 & 0.64 \\
 38 & 225.1311 &  36.4805 & 0.09256 &  1.45 & 225.1233 &  36.4833 & 225.1395 &  36.4775 &   1.2 &      62.8 & 0.93 \\
 39 & 225.2641 &  51.9671 & 0.13210 &  0.79 & 225.1911 &  51.9816 & 225.3009 &  51.9710 &   1.6 &      74.8 & 0.49 \\
 40 & 226.8555 &   8.4957 & 0.07867 &  0.77 & 226.8432 &   8.5091 & 226.8915 &   8.4469 &   1.8 &      77.3 & 0.30 \\
 41 & 227.4985 &   3.0010 & 0.09234 &  1.09 & 227.4847 &   2.9731 & 227.4993 &   3.0357 &   2.8 &      41.4 & 0.90 \\
 42 & 231.5492 &  29.6513 & 0.11664 &  1.79 & 231.5836 &  29.7171 & 231.5813 &  29.4925 &   1.2 &      65.4 & 0.45 \\
 43 & 238.0278 &  22.7943 & 0.11540 &  1.70 & 238.0168 &  22.8216 & 238.0665 &  22.6741 &   1.4 &      77.6 & 0.23 \\
 44 & 238.0383 &  20.0897 & 0.08950 &  2.08 & 238.1020 &  20.0863 & 237.8380 &  20.0530 &   0.8 &      15.3 & 0.31 \\
 45 & 238.0931 &  22.5533 & 0.06831 &  0.98 & 238.2262 &  22.6076 & 238.0541 &  22.5381 &   0.7 &      86.2 & 0.29 \\
 46 & 250.9306 &  32.5790 & 0.13525 &  1.17 & 250.8928 &  32.5756 & 250.9858 &  32.6105 &   0.6 &      75.6 & 0.57 \\
 \hline
\rule{0pt}{10pt} &       &       &      &    &       &  SRG     &       &       &    &        &   \\
 \hline
  1 & 120.4920 &  17.8646 & 0.14700 &  0.09 & 120.4966 &  17.8663 & 120.4876 &  17.8633 &   0.3 &      35.3 & 0.96 \\
  2 & 123.8014 &  38.6793 & 0.12500 &  0.08 & 123.8018 &  38.6830 & 123.8025 &  38.6739 &   1.2 &      81.3 & 0.68 \\
  3 & 125.6985 &   4.2972 & 0.09500 &  0.14 & 125.7078 &   4.2929 & 125.6884 &   4.3024 &   3.4 &      47.4 & 0.90 \\
  4 & 139.9519 &  57.8489 & 0.13700 &  0.37 & 139.9707 &  57.8497 & 139.9299 &  57.8540 &   1.2 &      12.3 & 0.79 \\
  5 & 141.5469 &  17.0661 & 0.11600 &  0.08 & 141.5501 &  17.0626 & 141.5431 &  17.0698 &   4.0 &      73.9 & 0.88 \\
  6 & 145.3501 &  39.7449 & 0.10800 &  0.16 & 145.3536 &  39.7554 & 145.3466 &  39.7352 &   2.0 &      53.7 & 0.93 \\
  7 & 145.5073 &   8.7936 & 0.13400 &  0.16 & 145.5155 &   8.7941 & 145.4975 &   8.7934 &   3.5 &      23.5 & 0.84 \\
  8 & 146.7625 &  23.2706 & 0.08400 &  0.11 & 146.7581 &  23.2783 & 146.7652 &  23.2618 &   0.3 &      34.7 & 0.95 \\
  9 & 153.9927 &  40.7798 & 0.12800 &  0.39 & 154.0132 &  40.7841 & 153.9673 &  40.7773 &   3.5 &      20.4 & 0.83 \\
 10 & 154.9770 &  39.5063 & 0.11200 &  0.25 & 154.9682 &  39.5185 & 154.9895 &  39.4932 &   1.2 &      38.2 & 0.86 \\
 11 & 155.4861 &  14.7254 & 0.11100 &  0.25 & 155.4970 &  14.7370 & 155.4747 &  14.7130 &   0.4 &      82.6 & 0.94 \\
 12 & 161.2614 &  47.2998 & 0.14500 &  0.09 & 161.2665 &  47.3000 & 161.2573 &  47.2993 &   0.4 &      49.2 & 0.82 \\
 13 & 161.9285 &  43.7813 & 0.08600 &  0.19 & 161.9416 &  43.7932 & 161.9162 &  43.7751 &   0.4 &      36.8 & 0.71 \\
 14 & 162.7186 &  40.0141 & 0.12900 &  0.50 & 162.7534 &  40.0142 & 162.6949 &  40.0138 &   0.7 &      41.9 & 0.68 \\
 15 & 165.1462 &  25.6531 & 0.14500 &  0.14 & 165.1428 &  25.6608 & 165.1494 &  25.6471 &   2.9 &      24.1 & 0.80 \\
 16 & 168.3246 &  41.4081 & 0.09500 &  0.31 & 168.3101 &  41.4158 & 168.3516 &  41.3934 &   1.5 &      63.9 & 0.53 \\
 17 & 172.8899 &  60.7911 & 0.14500 &  0.06 & 172.8910 &  60.7891 & 172.8865 &  60.7940 &   1.7 &      86.2 & 0.61 \\
 18 & 174.1098 &  50.2233 & 0.05400 &  0.08 & 174.1182 &  50.2169 & 174.1033 &  50.2301 &   0.8 &      22.3 & 0.96 \\
 19 & 176.1133 &  37.1422 & 0.11500 &  0.65 & 176.1444 &  37.1716 & 176.0805 &  37.1173 &   1.8 &      64.4 & 0.94 \\
 20 & 176.1171 &   3.9710 & 0.12700 &  0.08 & 176.1206 &   3.9717 & 176.1118 &   3.9693 &   2.5 &      36.6 & 0.64 \\
 21 & 177.4505 &  43.9035 & 0.07100 &  0.07 & 177.4534 &  43.9119 & 177.4491 &  43.8983 &   0.2 &      61.0 & 0.62 \\
 22 & 178.8550 &  25.5396 & 0.13700 &  0.08 & 178.8522 &  25.5427 & 178.8588 &  25.5360 &   4.3 &      59.8 & 0.81 \\
 23 & 181.8872 &  33.8778 & 0.07900 &  0.08 & 181.8805 &  33.8823 & 181.8923 &  33.8742 &   0.9 &      61.4 & 0.78 \\
 24 & 185.2185 &  31.5523 & 0.10400 &  0.06 & 185.2148 &  31.5553 & 185.2212 &  31.5504 &   1.5 &      39.7 & 0.69 \\
 25 & 198.7910 &   8.6815 & 0.09300 &  0.38 & 198.7985 &   8.7109 & 198.7845 &   8.6530 &   0.6 &      13.6 & 0.96 \\
 26 & 200.3242 &  42.5876 & 0.07900 &  0.34 & 200.3536 &  42.5874 & 200.2933 &  42.5799 &   2.3 &      60.2 & 0.90 \\
 27 & 207.8245 &  64.1603 & 0.10900 &  0.16 & 207.8355 &  64.1617 & 207.8143 &  64.1594 &   0.9 &      24.9 & 0.89 \\
 28 & 213.8634 &  17.4086 & 0.12400 &  0.06 & 213.8645 &  17.4124 & 213.8621 &  17.4057 &   4.4 &      25.0 & 0.82 \\
 29 & 222.4538 &  33.8574 & 0.08800 &  0.11 & 222.4477 &  33.8666 & 222.4574 &  33.8516 &   1.5 &      43.1 & 0.63 \\
 30 & 224.4742 &  28.5385 & 0.14400 &  0.56 & 224.4926 &  28.5609 & 224.4561 &  28.5140 &   0.2 &      85.8 & 0.94 \\
 31 & 227.1138 &  54.2521 & 0.09600 &  0.11 & 227.1230 &  54.2511 & 227.1070 &  54.2537 &   0.6 &      42.4 & 0.79 \\
 32 & 229.1675 &   0.2505 & 0.05300 &  0.26 & 229.1445 &   0.2715 & 229.1937 &   0.2267 &   1.4 &      71.4 & 0.88 \\
 33 & 230.6900 &  19.7056 & 0.10900 &  0.06 & 230.6949 &  19.7054 & 230.6871 &  19.7057 &   1.2 &      18.2 & 0.60 \\
 34 & 230.7391 &   2.9200 & 0.11000 &  0.10 & 230.7440 &   2.9155 & 230.7348 &   2.9243 &   1.3 &      25.5 & 0.90 \\
 35 & 239.7194 &  26.4386 & 0.08700 &  0.23 & 239.7098 &  26.4585 & 239.7263 &  26.4237 &   1.3 &      37.3 & 0.74 \\
\hline
\end{longtable}

\section{Uncertainty estimation for radio jet axis position angle} \label{sec:appendix-error}

To estimate the uncertainties in our measurements of the radio jet axis position angles, we adopt the standard formalism of \citet{Condon1997} and \citet{Fomalont1999}. For an unresolved or compact hotspot observed with a Gaussian restoring beam, the uncertainty in its centroid position is given by:
\begin{align*}
    \sigma_{\rm pos} \;\approx\; \frac{\mathrm{FWHM}}{2\,\mathrm{SNR}_{\rm peak}}
\end{align*}
where ${\rm FWHM}$  is the full width at half maximum of the beam, and ${\rm SNR}_{\rm peak}$ is the peak signal-to-noise ratio of the hotspot. In our calculations, we adopt ${\rm SNR}_{\rm peak}=3$ as a conservative lower limit when estimating centroid uncertainties.
Since the radio jet axis is defined by the line joining the two hotspots, the positional uncertainties of both hotspots contribute to the error in the measured position angle ($\sigma_{\rm \theta}$). Propagating these uncertainties yields:
\begin{align*}
    \sigma_{\rm \theta} \;\approx\; \frac{\sqrt{2}\sigma_{\rm pos}}{\mathrm{L}}\quad (\mathrm{radians})
\end{align*}
Here, $L$ (in arcsec) is the hotspot-hotspot separation, i.e., the projected angular size of the GRG. Using the median angular size of our sample ($L = 9.67\arcmin$) and adopting ${\rm SNR}_{\rm peak}=3$, we obtain position-angle uncertainties of $\sigma_{\theta}\approx 0.93^{\circ}$ for the LOFAR low-resolution images (144\,MHz, FWHM $= 20\arcsec$) and $\sigma_{\theta}\approx 0.28^{\circ}$ for the LOFAR high-resolution images (144\,MHz, FWHM $= 6\arcsec$). For the FIRST (1400\,MHz, FWHM $= 5\arcsec$) and NVSS (1400\,MHz, FWHM $= 45\arcsec$) images, the corresponding uncertainties are approximately $0.23^{\circ}$ and $2^{\circ}$, respectively. Thus, the radio jet axis uncertainties in our measurements are typically well below $1^{\circ}$, and are far too small to influence the statistical results or the interpretation of the orientation analysis.

\end{appendix}

\end{document}